\documentstyle [12pt,epsfig]{article} 
\makeatletter

\@addtoreset{equation}{section}
\makeatother
\renewcommand{\theequation}{\thesection.\arabic{equation}}
\newcommand{\ba}{\begin{eqnarray}}
\newcommand{\ea}{\end{eqnarray}}

\begin{document}
\newcommand{\BS}{\bigskip}
\newcommand{\SECTION}[1]{\BS{\large\section{\bf #1}}}
\newcommand{\SUBSECTION}[1]{\BS{\large\subsection{\bf #1}}}
\newcommand{\SUBSUBSECTION}[1]{\BS{\large\subsubsection{\bf #1}}}

\begin{titlepage}
\begin{center}
\vspace*{2cm}
{\large \bf The physics of space and time I: The description of rulers and clocks in
 uniform translational motion by Galilean or Lorentz transformations}  
\vspace*{1.5cm}
\end{center}
\begin{center}
{\bf J.H.Field }
\end{center}
\begin{center}
{ 
D\'{e}partement de Physique Nucl\'{e}aire et Corpusculaire
 Universit\'{e} de Gen\`{e}ve . 24, quai Ernest-Ansermet
 CH-1211 Gen\`{e}ve 4.
}
\newline
\newline
   E-mail: john.field@cern.ch
\end{center}
\vspace*{2cm}
\begin{abstract}
  A calculus based on pointer-mark coincidences is proposed to define, in a mathematically
  rigorous way, measurements of space and time intervals. The connection between such
   measurements in different inertial frames according to the Galilean or Lorentz 
   transformations is then studied using new, simple, and practical clock synchronisation
   procedures. It is found that measured length intervals are Lorentz invariant, whereas moving 
  clocks show a universal time dilation effect. The `relativistic length contraction'
   and `relativity of simultaneity' effects of conventional special relativity theory are shown
  to be the consequence of calculational error. Two derivations of the Lorentz transformation
  from simple postulates, without reference to electrodynamics or any other dynamical theory,
   are reviewed in an appendix.

\vspace*{1cm}
\end{abstract}
\end{titlepage}
 
\SECTION{\bf{Introduction}}
    The historical development of Special Relativity (SR), which may be defined as the 
    physical theory of space and time, in the absence of gravitational effects, was intimately
    connected during the second half of the 19th Century with that of classical electrodynamics.
    The transformation of time and of transverse
    spatial coordinates was proposed by Voigt~\cite{Voigt} in order
    to leave invariant the form of the wave equation in different
    inertial frames. Subsequently, first Larmor~\cite{Larmor}, and later Lorentz proposed~\cite{Lor}
     a similar transformation of time, as well as 
    a transformation of the longitudinal spatial coordinate, to obtain a similar invariance
    of the free-space Maxwell Equations\footnote{It is difficult to understand why Larmor's
    work, published in 1900, was not cited in Lorentz' 1904 paper where identical
    transformation equations were given. See Ref.~\cite{CK} for
     a discussion of priority issues related to the discovery of the Lorentz Transformation.}.
 This close connection has persisted in text-book
    presentations of SR until the present day. An example is the emphasis that is still placed,
     following Einstein's seminal paper~\cite{Ein1}, on the space-time properties of
      `light signals', `light waves' and, in particular Einstein's light-signal synchronisation
    procedure, when deriving or discussing the space-time Lorentz Transformation (LT).
    The latter, with the aid of the further definitions of relativistic velocity, momentum and
    energy (see Section 7 below) enable all the kinematical predictions of SR to be derived.
    \par The aim of the present paper is to present the space-time LT ---which is the essence
    of SR--- in a completely new light, without any reference to electrodynamics or any
     other dynamical theory. In particular, two aspects are discussed.  The first is the axiomatic
    basis of the LT. This is done, in a general way, below, in the present section, 
     where some sets of postulates, from which the LT may be derived, are presented and 
    motivated, and in more detail in the Appendix where two previously
    published~\cite{JHF1,JHF2} derivations of the LT are reviewed. The second and more novel
   aspect of this
   paper is a careful application of the Galilean Transformation (GT) or the LT
   to measurements of space and time intervals performed, with the aid of rulers and clocks,
    in different inertial frames. 
   Important for this discussion are, firstly, the method by which space and time intervals
    are experimentally determined, and secondly, synchronisation procedures for spatially 
    separated clocks in the same, or different, inertial frames. The procedures proposed
   avoid completely the `conventionality'~\cite{Convt} associated with the Einstein
    light signal method~\cite{Ein1} and the subsequent necessity of postulating, in SR, the
    spatial isotropy of the speed of light. 
   \par The present study provides a simple and clear explanation for the shortcomings
    in the current physical interpretation of the space-time LT that have been pointed out, from
    a different point-of-view in a previous paper~\cite{JHF3} by the present author. 
    In fact the `relativity of simultaneity' and `length contraction' effects of SR, for which,
    unlike time dilation, no experimental evidence exists~\cite{JHF3}, are found to be illusory
    ---the consequence of a trivial mathematical error in the description of synchronised 
     clocks by the LT equations.
    \par Since the first version of the present paper was completed, but not released, 
      in late 2005, a number of short papers have been written by the present author
     explaining in either a logically concise~\cite{JHFUMC} or in a
      pedagogical~\cite{JHFCRCS,JHFACOORD} manner the conclusions
     of Ref.~\cite{JHF3} and the present paper.
      \par Some postulates which may be used to derive the LT  are now presented and
   briefly discussed. In his original special relativity
   paper~\cite{Ein1}, Einstein explicitly stated\footnote{There are also several other postulates
   that are tacitly assumed, such as linearity of the equations, spatial isotropy and the Reciprocity
   Principle~\cite{BG} which is discussed in the Appendix of the present paper.
    For more details see Reference~\cite{JHF1}.}
  two postulates, (E1) and (E2) from which he proceeded to his derivation of the space-time
  LT. These were:
 \par {\bf (E1) The dynamical laws of nature are the same in any inertial frame.}
 \par {\bf (E2) The speed of light is the same in any inertial frame, and is independent
   of whether the source is stationary or in motion\footnote{Actually, Einstein did not explicitly
    state in his second postulate that the speed of light is the same in all inertial frames, 
    only that it is independent of the motion of its source. The constancy of the 
     speed of light was considered to be a `law of nature', and so a particular case of the postulate E1.}.} 
  \par The postulate (E1), the `Special Relativity Principle', was not new. Indeed it is equally
    valid in pre-relativistic Newtonian mechanics, and had been clearly stated~\cite{Gal} already
   by Galileo. The second postulate, (E2), is highly counter-intuitive. Common sense 
   would suggest that if light is of a corpuscular nature, as imagined by Newton, the speed of
   a particle of light would be expected to be greater, if the source moves in the
   same direction as its motion, and smaller, if it moves in the opposite direction, than
   light emitted from a source at rest. If, on the other hand, it is supposed that light is some 
    kind of wave motion, as was generally assumed in the second half of the 19th Century, it is
    expected that some kind of prefered frame will exist ---that of the medium supporting the
     wave motion that is identified with light--- the `luminferous aether'. In this case the
    speed of light is expected to be different for observers with different velocities
     relative to the aether frame.
    \par However, it was very soon realised that the postulate (E2) is not necessary to derive
    the LT. Indeed, already as early as 1910, Ignatowsky~\cite{Ignatowsky} published a derivation of the LT
    which (although the postulates on which it is based are not explicity stated) is similar,
    in its essential features, to the first derivation of the LT presented in the Appendix
    of the present paper. That the postulate (E2) is not necessary to derive the LT
     was pointed out by Pauli in the monograph on relativity~\cite{Pauli} that he
     wrote in 1921. At the time of this writing, the literature on derivations
   of the LT {\it not} using postulate (E2) is vast. A partial list up until 1968 can be
   found in a paper by Berzi and Gorini~\cite{BG}. A survey of the more recent literature
   is given in the paper~\cite{JHF1} by the present author.
    \par These `lightless' derivations of the LT can be divided into two broad categories:
    \begin{itemize}
     \item[(i)] Those in which the postulate (E2) is replaced by some other `strong' 
        kinematical consequence of the LT such as conservation of relativistic transverse
       momentum or relativistic `mass increase'.
     \item[(ii)] Those in which the minimum number of, and only very weak, postulates
                 are employed.
    \end{itemize}
      Examples of derivations of type (ii) may be found in Refs.~\cite{LK,LLB,Mermin,AS} and 
      Refs.~\cite{JHF1,JHF2} by the present
     author. For the reader's convenience, the essential features of two latter derivations are 
     recalled in the Appendix of the present paper. The postulates used in these derivations
     are the following:
     \par {\bf (A) Each Lorentz Transformation equation must be a single-valued function
               of all its arguments.}
      \par {\bf (B) Reciprocal space-time measurements of similar rulers and clocks 
            at rest in two different inertial frames S, S', by observers at rest in S', S 
             respectively, yield identical results.}  
       \par {\bf (C) The equations describing the laws of physics are invariant with respect
             to the exchange of space and time coordinates or, more generally, with respect
            to the exchange of spatial and temporal components of 4-vectors.}
      \par The postulate (B) is a generalisation of Pauli's postulate (c)\footnote{This postulate
      states that, in the notation of the present paper, ``the contraction of lengths at rest in
       S' and observed in S is equal to the contraction of lengths at rest in S and observed in S'.''.}
      ~\cite{Pauli}.
      \par The derivation of Ref.~\cite{JHF1} is based on (A) and (B), that of Ref.~\cite{JHF2}
        on (A) and (C). These postulates alone are sufficient to derive the LT for space
          and time intervals along the common $x$, $x'$ axis of two inertial frames S and
           S' in relative motion parallel to this axis. The additional postulate of spatial 
      isotropy is needed for the transformation of events lying outside
       the  $x$, $x'$ axis.
      \par It is interesting to compare the postulates  (A), (B) and (C) with the 
       Einstein postulates (E1) and (E2). (A) and (C) are essentially mathematical
      statements about the structure of physical equations but make no reference to any dynamical
     or kinematical laws of physics. 
        \par The meaning of postulate (A) is that, if the LT equations are written
             as
           \[f_x(x',x,\tau,v) = 0,~~~ f_t(t',x,\tau,v) = 0 \]
         where $\tau$ and $t'$ are times
        recorded by observers at rest in S of clocks at rest in S and S' respectively, and
        $v$ is the velocity of S' relative to S along the $x$, $x'$ axis, the solution
       for $\alpha_1$, given $\alpha_2$ 
       and $\alpha_3$ must be a single valued function of  $\alpha_2$ and $\alpha_3$.
        Here, $\alpha_1$  $\alpha_2$ and $\alpha_3$  are any permutation of $x'$, $x$ and $\tau$
        for $f_x$ and of  $t'$, $x$ and $\tau$ for $f_t$. A sufficient condition for this is that
         $f_x$ and  $f_t$ be multilinear functions of their arguments. The physical 
        basis of the postulate (A) is very obvious: One, and only one, event ($x'$, $t'$)
        can exist in S' for each event ($x$, $\tau$) in S.
          \par The postulate (C) makes a much more specific and powerful statement about
       the structure of physical equations but makes no reference, unlike (B), to inertial
       frames or the results of space-time measurements performed in these frames.
        Still, in combination with the weaker (but still purely mathematical) postulate (A)
        it is sufficient to derive the LT.
       \par The postulate (B), which was termed the `Kinematic Special Relativity Postulate' or
        KSRP in Refs.\cite{JHF1,JHF2}, but would be more properly called the `Measurement
        Reciprocity Postulate', or MRP, may seem obvious to some readers. Indeed, it is often
       tacitly assumed to hold in derivations of the LT. This was effectively done in Einstein's
       original derivation. However, as will be demonstrated below, only kinematical definitions
       and pure logic are needed to show that Einstein's counter-intuitive postulate (E2)
       is a necessary consequence of the `obvious' postulates (A) and (B).
         \par All of the physical consequences of SR can be derived from the space-time LT,
         either directly, by applying them to space or time measurements performed in different
     inertial frames, as discussed in detail below in the present paper, or by using them to
     define the transformation properties of other quantities of physical interest.
     However, to {\it derive} the LT only the weak and `self apparent' postulates (A) and (B)
     are necessary.
         \par Given the title of Einstein's original SR paper `On the  Electrodynamics of Moving
      Bodies' and that of Poincar\'{e}'s SR paper~\cite{Poin} published in the following year
     `On the Dynamics of the Electron' what then is the essential connection between 
      electrodynamics and the fundamental physics of SR? The perhaps surprising, but simple,
     answer is: `None'. The properties of space and time (actually because the subject here is
     physics, those of {\it measurements} of space and time) underlie and take precedence
     over any dynamical aspect of physics, with the possible exception of the general
     relativistic theory of gravitation, which is not considered here.
     These properties may be compared to a canvas on
    which the electromagnetic, weak and strong interactions paint the picture that we perceive
    as the physical world. Paint is not needed to make canvas. The subject of this paper
    is just the canvas, that is, the physics of space and time intervals as measured
    in different inertial frames in the absence of gravitational effects.
     \par Spatial intervals are measured by rulers and temporal
    intervals by clocks. Therefore the subject of the present paper reduces essentially to a study
    of the {\it modus operandi} of rulers and clocks as measuring instruments. The important
    problem of synchronisation of spatially-separated clocks is addressed from scratch.
    A simple procedure of master clock synchronisation (synchronisation by `pointer transport')
    is suggested to synchronise an arbitary number of spatially separated clocks in
    a given inertial frame. An equally simple method (synchronisation by `length transport')
    that may be used to synchronise an arbitary number of spatially separated clocks in the
    same, or different, inertial frames is also proposed. When this is done, it is seen that 
    measurements of the space intervals between stationary objects, or of objects moving
    with the same velocity, are the same whether the GT or the LT is used. The measured spatial
     separation
    of any two such objects, or equivalently the length of a ruler, is a Lorentz invariant quantity
    ---it is independent of the frame in which it is measured. There is no relativistic
    `length contraction'. Uniformly moving clocks all appear to run slow according to
    the time dilation factor $\sqrt{1-(v/c)^2}$, but all synchronised clocks in 
    any inertial frame show the same time, as viewed from any other inertial frame
     ---there is no `relativity of simultaneity' in this case.
     \par The structure of this paper is as follows: In the following section the measurement
    of space and time intervals is discussed at a fundamental level, where each measurement
    is identified with a spatial pointer-mark coincidence. The equivalence of the physical
    concept of time with that of uniform motion, or cyclic motion of constant period, is
    stressed. In Sections 3 and 4 the the pointer transport and length transport
    synchronisation methods, respectively, are presented. In Section 5 the relation of
    space and time measurements in different inertial frames is considered. The different
    predictions of the GT and the LT are compared and contrasted. Section 6 explains
    the spurious and unphysical nature of the related `relativity of simultaneity' and
    `relativistic length contraction' effects of conventional special relativity
    theory. It is shown in Section 7 how Einstein's
    second postulate (E2) is a consequence of indentifying light with massless particles
    ---photons. Section 8 contains a summary and a brief discussion of the impact of the
     conclusions of the present paper and Ref.~\cite{JHF3} on other aspects of SR.

\SECTION{\bf{The measurement of space and time intervals}}

\begin{figure}[htbp]
\begin{center}\hspace*{-0.5cm}\mbox{
\epsfysize15.0cm\epsffile{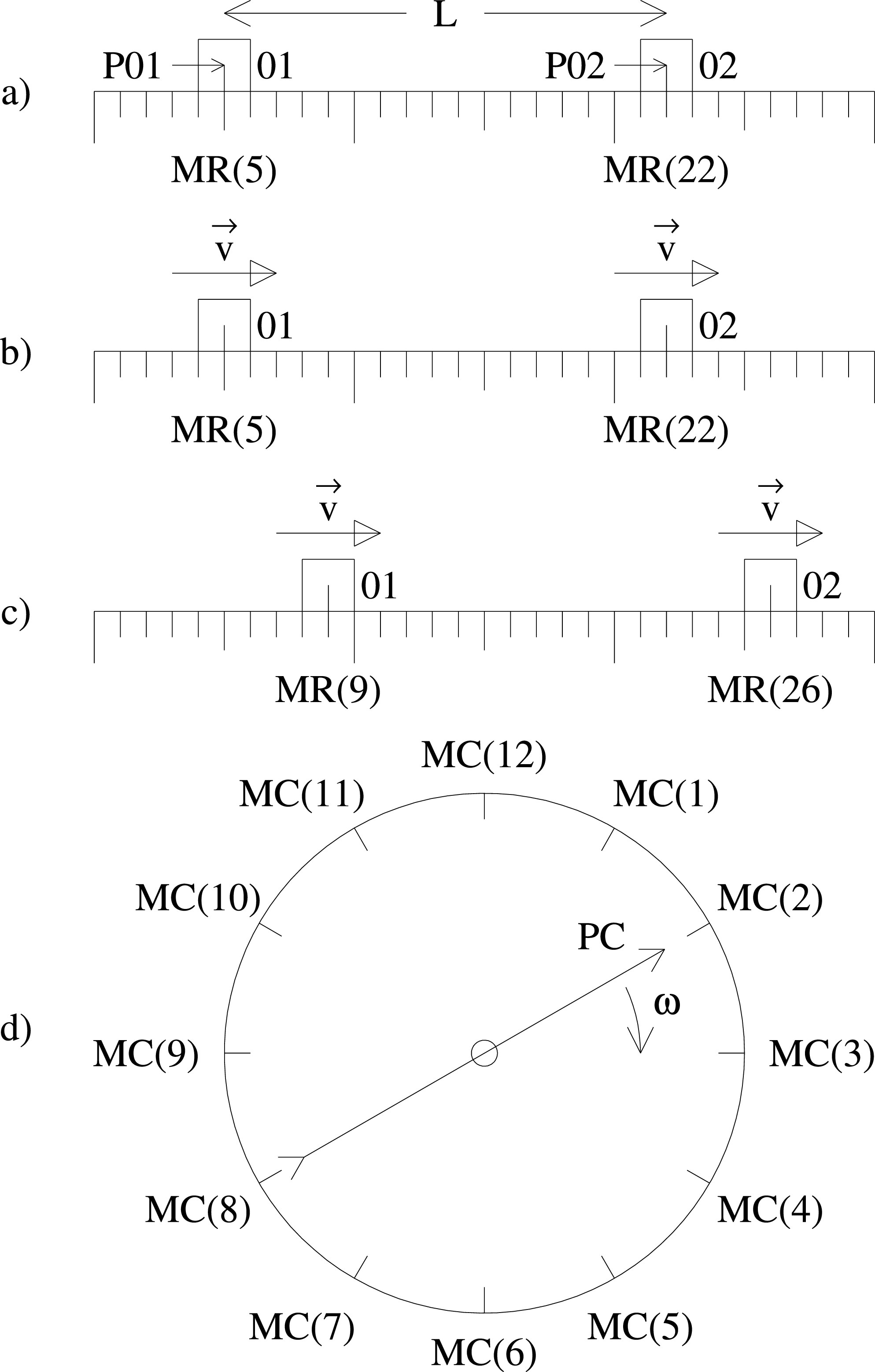}}
\caption{ {\em a) Measurement of the distance,L, between two objects O1,O2, at rest, by a ruler.
  The pointers $PO1$, $PO2$ on the objects are in spatial coincidence with the ruler marks
  $MR(5)$ and $MR(22)$. The times of the corresponding PMC: $P01@MR(5)$ and $P02@MR(22)$
  are arbitary. b),c) measurements of the distance between the same two objects when they are both 
   moving parallel to the ruler with the same uniform velocity, $v$. The corresponding PMCs:
  $P01@MR(5)$ and $P02@MR(22)$ in b) and $P01@MR(9)$ and $P02@MR(26)$ in c) must be simultaneous
  in each case.
  d) Time measurement by an analogue clock in terms of the PMC between the hand, constituting
   a time-dependent pointer $PC$, and the uniformly spaced marks $MC(1)-MC(12)$. The hand rotates
   with constant angular velocity $\omega$.}}
\label{fig-fig1}
\end{center}
\end{figure}

  The equations of physics contain symbols representing quantities at many
  different levels of mathematical abstraction from those representing the raw data of any 
  experiment. However, in order to provide meaningful predictions for the
  result of any experiment, the mathematical expression of any physical
  theory must contain, at least, symbols in exact correspondence with
  experimental raw data. Only in this way is it possible to compare
  theory and experiment. Particularly for space and time measurements, great care
  must be taken to ensure this exact correspondence between raw data and the
  theoretical symbols that represent them. The raw data itself however has a very simple
  and universal form. As succinctly stated in Ref.~\cite{RBLHM}:
  \par {\tt However, every experiment in physics in which measurement plays a part
   \newline reduces
  essentially to an operation involving the observation of the \newline coincidence of a point
   or pointer with a mark on some scale, or the\newline comparison of sets of such
   coincidences; and
   the symbolic expression of the observation is made by means of numbers associated 
   with the marks. }
  \par This procedure evidently applies to the measurement of the length of an object or
    the distance between two objects, but, as will be seen, it applies equally to the
    measurement of time intervals ---that is the raw data of time measurements can always
   be expressed in terms of the measurement of some corresponding spatial interval. 
   For time measurements the concept of local Pointer-Mark Coincidences, denoted in the
   following, for brevity, as PMC, is therefore equally important. 
   \par In Einstein's first paper on special relativity~\cite{Ein1} can be found the
     following important statement concerning the measurement of time:
   \par  {\tt We have to take into account that all our judgements in which time
     plays a part are always judgements of} {\it simultaneous events}. {\tt If, for
     instance, I say ``That train arrives here at 7 o'clock'', I mean
     something like this: ``The pointing of the small hand of my watch to 7 and the
     arrival of the train are simultaneous events''.}
    \par Observation of the time of arrival of the train then requires the simultaneous
       occurence of the two following PMC:
 \begin{itemize}

     \item[(i)] Of a pointer on the train with a mark on the platform
                at the same position as the observer with the watch.
    \item[(ii)] Of the pointer constituted by the small hand of the watch
                with the mark correponding to `7 hours' on the dial
                  of the watch.
 \end{itemize} 
    \par In order to specify, in a precise mathematical manner, the raw
   data of space and time measurements in the following (which definitions
   must be rigorously respected by the symbols of the theory of space and 
   time) a symbolism for pointers, marks and pointer-mark coincidences is now 
   introduced. The spatial coincidence of a pointer,  P, with a mark, M, at time, t,
   defines a PMC at time t : $PMC(t) \equiv P(t)@ M$. As indicated, pointers are 
   usually (but not necessarily) time dependent whereas marks are time-independent
   and in one-to-one coincidence, in an obvious way, with numbers that represent Cartesian
   or angular coordinates. Simple examples are the marks on a ruler or on the
    dial of an analogue clock. 
    \par Einstein's measurement of the arrival of a train at 7 o'clock at Bern station
    is then expressed symbolically as follows:
    \[ PMCT(t_T) \equiv PT(t_T) @ MP,~~~PMCWH(t_T) \equiv PWH(t_T) @ MWH(7), \]
     \[ PMCWM(t_T) \equiv PWM(t_T) @ MWM(0), \]
   \[ x_T = x(MP),~~~ t_T = t(MWH(7),MWM(0)). \]
    Here $PT$, $PWH$ and $PWM$ are a pointer (P) on the train (T) and the hour (H) and minute (M) hands of
  Einstein's watch respectively. $MP$ is a mark (M) on the station platform (P) at Bern and 
  $MWH(7)$, $MWM(0)$, are the marks for 7 hours and 0 minutes on the watch dial.
  If Einstein's watch is correctly synchronised with the Bern master clock
  then $t_T =$ 7h00min Bern time. If not, the functional dependence of
   $ t_T = t(MWH(7),MWM(0))$ is more complicated. If Einstein's watch is in advance
     or retarded relative to the Bern master clock additional constants must be
     respectively subtracted from or added to $t_T$. A suitable synchronisation procedure
    based on a master clock is described in the following section.
  \par Some examples of elementary space and time measurements are shown in Fig.~1. 
   In Fig.~1a, the spatial separation between two objects O1 and O2 at rest is measured
   via the marks (M) $MR(J)$ on a ruler (R). Here $J$ is the ordinal number of the ruler mark.
   To each mark is associated a number specifying a Cartesian coordinate: $x_J = x[MR(J)]$
   The PMC specifying this measurement are:
   \begin{equation}
    PMC1(t_1) \equiv PO1(t_1) @ MR(5),~~~ PMC2(t_2) \equiv PO2(t_2) @ MR(22).
   \end{equation}
  The measured separation of the objects is then:
   \begin{equation}
   L \equiv L(0) =  x[MR(22)]-x[MR(5)].
   \end{equation}
   $L(v)$ is the spatial separation  of the objects when they are both moving
    with velocity $v$ parallel to the ruler. It may be noted  that in this case,
    since the objects are at rest, the times $t_1$ and $t_2$ at which $PMC1$ and
    $PMC2$ are established are arbitary. 
    \par In Fig.~1b a similar measurement of the spatial separation of O1 and O2
    is made except that both objects now move with uniform speed $v$ parallel to the
    ruler. Such a measurement {\it requires simultaneity} of $PMC1$ and $PMC2$:
   \begin{equation}
    PMC1(t_1) \equiv PO1(t_1) @ MR(5),~~~ PMC2(t_1) \equiv PO2(t_1) @ MR(22).
   \end{equation}
  The measured separation, $L(v,t_1)$ is still given by Eq.~(2.2).
  The measurement may be performed at any other time provided that the condition
  of simultaneity is respected. For example, as shown in Fig.~1c:
   \begin{equation}
    PMC1(t_2) \equiv PO1(t_2) @ MR(9),~~~ PMC2(t_2) \equiv PO2(t_2) @ MR(26)
   \end{equation}
  while
   \begin{equation}
    L(v,t_2) =  x[MR(26)]-x[MR(9)].
   \end{equation} 
 It is clear from the uniformity of the marks on the ruler that all three
  measurements of the spatial separation of the objects give the same result:
  \begin{equation}
    L \equiv L(0) =  L(v,t_1) =  L(v,t_2).
   \end{equation} 
  The measurement of the spatial separation of two objects with equal
  uniform motion then evidently, as in the case of the time measurement
  mentioned by Einstein and quoted above, requires {\it simultaneous} observation of two
  distinct $PMC$. The simultaneity requirement thus presupposes the existence
  of synchronised clocks at the spatial locations of the two $PMC$. In Section 4
  below this relation will be inverted to provide a method to synchronise clocks
  in the same, or different, inertial frames.
  \par The spatial configurations of the objects and the ruler in Fig.~1a,b,c 
    provide independent and consistent measurements of the spatial separation of the
  objects, whether they are at rest or in uniform motion. The comparison of the
  configurations in Fig.~1b and 1c also provides the simplest possible example of
  the measurement of a time interval. For this it is necessary to associate
 $PMC1(t_1)$ with $PMC1(t_2)$ or $PMC2(t_1)$ with $PMC2(t_2)$. That is to consider the
 coincidences of a given pointer with two different marks\footnote{ That measurements of
  length intervals require $PMC$ between different pointers and different marks, whereas
 time interval measurements require $PMC$ between the {\it same} pointer and the same, or
 different, marks was previously pointed out by G. Burnison-Brown~\cite{GBB}.}
    \begin{equation}
    PMC1(t_1) \equiv PO1(t_1) @ MR(5),~~~ PMC1(t_2) \equiv PO1(t_2) @ MR(9),
   \end{equation} 
  \begin{equation}
   \Delta t(1) = (t_2-t_1)_1 = \frac{1}{v}\{ x[MR(9)]- x[MR(5)]\}
   \end{equation} 
or
     \begin{equation}
    PMC2(t_1) \equiv PO2(t_1) @ MR(22),~~~ PMC2(t_2) \equiv PO2(t_2) @ MR(26),
   \end{equation} 
  \begin{equation}
   \Delta t(2) = (t_2-t_1)_2 = \frac{1}{v}\{ x[MR(26)]- x[MR(22)]\}.
   \end{equation} 
  From the uniformity of the ruler marks it is clear that:
  \begin{equation}
   \Delta t(1) =  \Delta t(2) \equiv  \Delta t.
   \end{equation} 
   In this example it is evident that the concept of time is inseparable from that
 of rectilinear motion with the {\it uniform speed} $v$. Comparing Fig1c and 1d it can be
 seen that the measurement of a time interval by an analogue clock is in everyway
 similar to the example just discussed. The ruler in Fig1c is bent into the form of a circle
 and the objects O1 and O2 moving with uniform speed in a straight line  are replaced by the hand
  of the clock, $PC$,
 rotating with uniform angular velocity $\omega$.
  \par An alternative way to introduce the concept of a measurable time into physics 
    is to assume the existence of physical systems with a reproducible periodic motion.
 In this case, the concept of a measurable time interval is inseparable from that of the
 existence of physical systems executing cyclic
  motion with a {\it constant period}. It is then clear, firstly, that `time' and 
  `uniform motion' are inseparable concepts in physics and, secondly, that time measurements
   may always be constructed from $PMC$, that is from {\it spatial} pointer-mark coincidences,
   given the existence of physical systems executing uniform motion, whether rectilinear 
   or periodic. For the case of periodic motion, the $PMCs$ must contain the {\it same} pointer
   and mark. Consider, for example, the analogue clock of Fig.~1c where the 
  pointer (hand) $PC$ rotates with constant angular velocity $\omega$. The period is defined
  by successive $PMC$ of the type $PC@MC(J)$. The corresponding period is $\tau = 2 \pi/ \omega$.
 
\SECTION{\bf{Clock synchronisation by pointer transport}}

\begin{figure}[htbp]
\begin{center}\hspace*{-0.5cm}\mbox{
\epsfysize15.0cm\epsffile{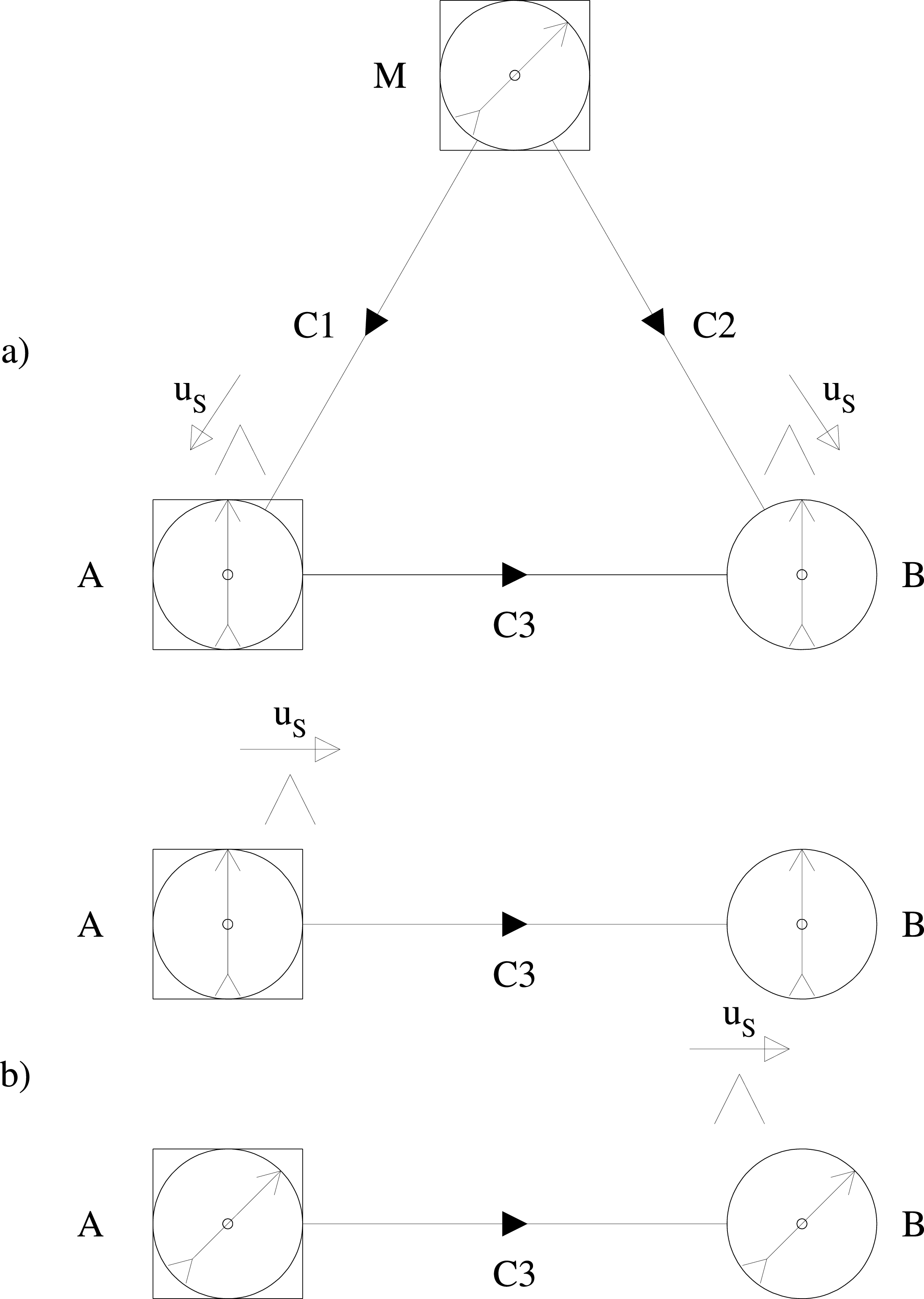}}
\caption{ {\em a) A procedure to synchronise two clocks A and B in the same inertial frame
 by pointer transport using a master clock M (see text). b) Measurement of the signal speed
 $u_S$ in the cable C3, of known length, using the synchronised clocks A and B.}}
\label{fig-fig2}
\end{center}
\end{figure}

\begin{figure}[htbp]
\begin{center}\hspace*{-0.5cm}\mbox{
\epsfysize10.0cm\epsffile{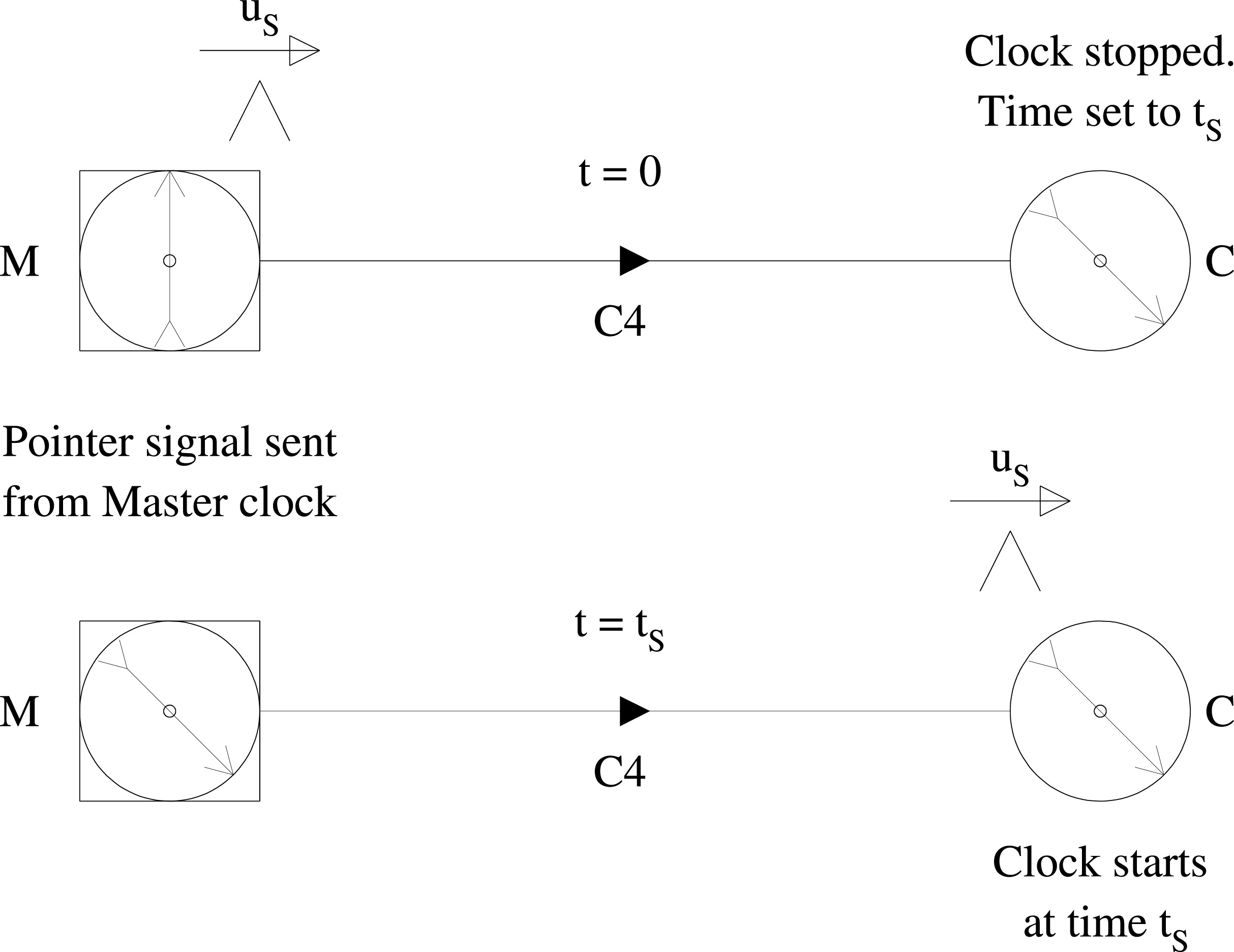}}
\caption{ {\em A procedure to synchronise a clock at an arbitary position in the same frame
   as the master clock M. The cable C4 is of known length, $\ell_4$ and known signal speed $u_S$.
     When the clock C starts at the time $t_S = \ell_4/u_S$ it is synchronous with M. }}
\label{fig-fig3}
\end{center}
\end{figure}
   As described in the previous section, in order to measure the length interval between
  two moving objects, two instantaneous $PMC$ must be recorded. This requires, in turn, that
  synchronised clocks are placed at each ruler mark. The present section describes how an 
  arbitary number of clocks, all at rest in a common reference frame, may be synchronised with
  a master clock, M, also at rest in the same frame. 
  \par It is assumed that all clocks may be connected to M by cables C1,C2,... of identical 
   construction but of different lengths, so as to conveniently link all of the 
   to-be-synchronised clocks to M. The pointer is an electrical signal generated at a known value
   of the time recorded by the master clock. The slave clocks are stopped at the beginning of the
   synchronisation procedure and are started by the PMC corresponding to receipt of the
    pointer signal from M. One other clock, A, is also equipped with a pointer signal generator
   similar to that of M. This clock may then be used as a secondary master clock. Three
   steps are necessary to synchronise an arbitary clock, C, at some fixed position in
   the same reference frame, with M:
  
  \begin{itemize}
    \item[(a)] The clocks M, A and B, assumed all to run at the same rate, are connected 
               as shown in Fig.~2a. Cables C1 and C2 are of equal, but arbitary, length (it
                 does no need to be known) whereas C3 is of precisely known length.
                Simultaneous pointer signals are sent from M along C1 and C2 to A and B.
                The latter two clocks are initially stopped and  set to time zero.
                 On receipt of the pointer signals A and B start and thereafter indicate
                synchronous times. The start signals are received simultaneously
                by A and B because the cables C1 and C2 are of equal length, and being of
               similar construction have the same signal propagation velocity $u_S$.

   \item[(b)]  As shown in Fig.~2b, a pointer signal is now sent, at known time $t_A$,
               from A to B along C3, after these clocks have been synchronised as described
               in (a). Reception of the signal stops B at the time $t_B$. If $\ell_3$ is the
                length of cable C3 (suitably corrected to take into account signal 
                 transit times in A and B) the speed of signal propagation is measured to
                be: 
                  \begin{equation}
                   u_S = \frac{\ell_3}{t_B-t_A}.   
                 \end{equation}

    \item[(c)]  In order to synchronise with M an arbitary clock C, at any distance from M,
                it is connected, to M by a cable, C4, of arbitary but
                 known length, $\ell_4$.
                 This cable is of similar construction to C1, C2 and C3 and has the same signal
                 propagation speed $u_S$. The clock C is stopped and set to the time
                 $t = t_S = t_0+\ell_4/u_S$. where $t_0$ is a  convenient, pre-determined, time offset, and
                 $u_S$ has been determined as in (b) above.
                 At time $t = t_0$ a pointer signal is sent from M to C. When the clock C is started
                 by the pointer signal at M time $t_S$ it is synchronous with M. This synchronisation
                 procedure is illustrated In Fig.~3 for the case $t_0 = 0$.
   \end{itemize}
    
            \par The above method enables synchronised clocks to be placed at arbitary 
                 positions in any given frame of reference, but does not address the 
                 problem of synchronisation of clocks in frames in relative motion.
                  A method to synchronise an arbitary number of clocks in an inertial
                 frame with a similar array of clocks in another inertial frame 
                 is described in the following section.
 
\SECTION{\bf{Similar motion of two objects in a single inertial frame. Clock synchronisation
   by length transport. The moving ruler clock}}

\begin{figure}[htbp]
\begin{center}\hspace*{-0.5cm}\mbox{
\epsfysize10.0cm\epsffile{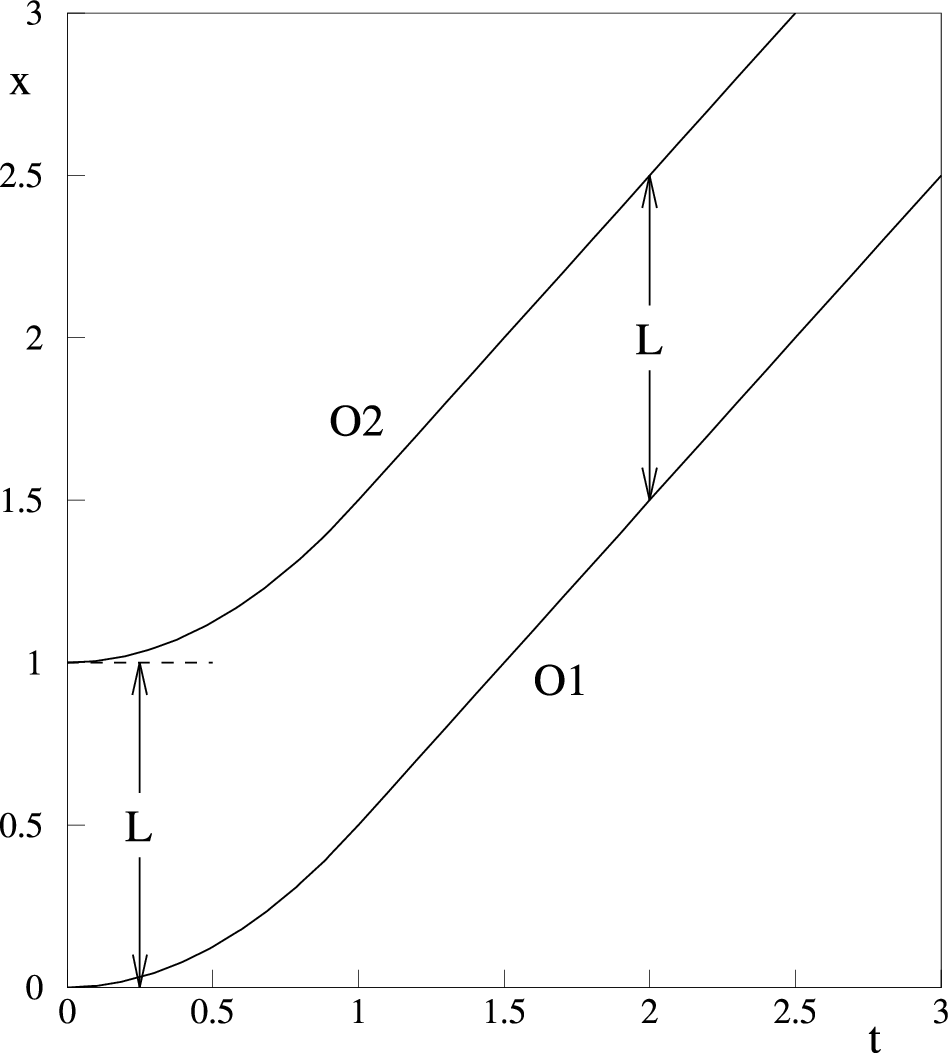}}
\caption{ {\em Space-time trajectories of two objects O1 and O2 undergoing equal uniform
 acceleration during one unit of time in the inertial frame S. The distance between the 
 objects is $L$ at all times. The distance between the objects is also $L$ at all times in their 
 common, instantaneous, co-moving inertial frame, S'. As the two objects are referred, in both cases,
  to a single frame of reference, the above statements concerning their separation are valid
  for a space-time geometry described by either the Galilean or the Lorentz transformation.
  These transformations relate only space and time measurements, performed in different inertial frames,
  of a single object.}}
\label{fig-fig4}
\end{center}
\end{figure}

\begin{figure}[htbp]
\begin{center}\hspace*{-0.5cm}\mbox{
\epsfysize15.0cm\epsffile{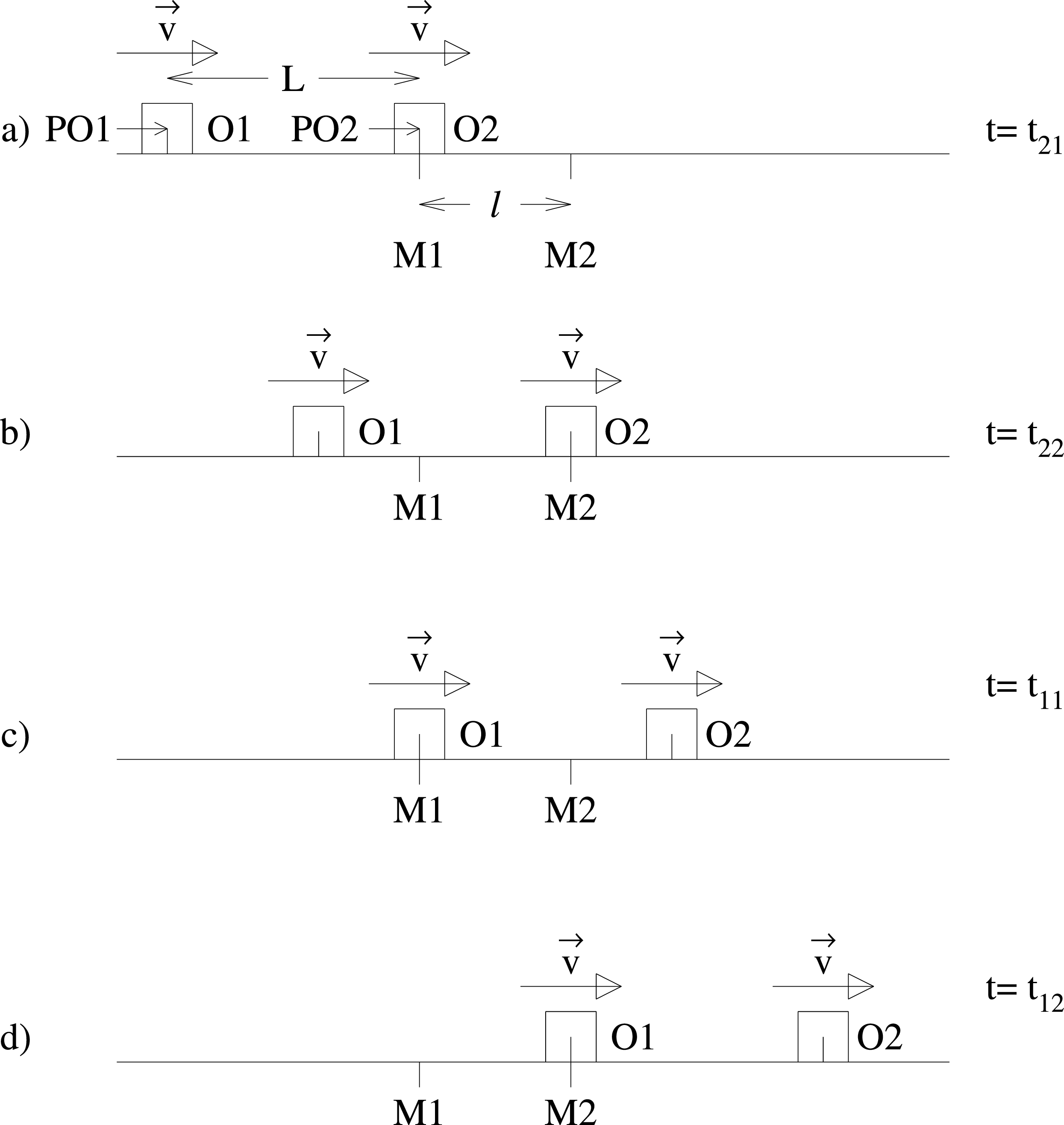}}
\caption{ {\em Definitions of the different PMC used to measure the spatial separation
  of two objects O1 and O2 moving parallel to the ruler $M1-M2$. See Eq.~s(4.5)-(4.10).}}
\label{fig-fig5}
\end{center}
\end{figure}

    The motion along the x-axis of an inertial frame S of two objects O1 and O2 is now considered.
    At time $t = 0$ the objects are at rest and separated by the distance $L$. Each object is
    then subjected to an identical acceleration program during the time $t_{acc}$ in S. The 
    instantaneous co-moving inertial frame of the objects, at any instant, is denoted by S'. For $t > t_{acc}$,
    S' is a fixed inertial frame moving with speed  $v$ relative to S. For definiteness the acceleration
    program is chosen to correspond to a constant acceleration, $a$, in the frame S during the 
    period $0 < t < t_{acc}$. In this case the positions of the objects O1, (O2) in S: $x_1(t)$, ($x_2(t)$)
    at time $t$ are given by the equations:
     \[0 < t < t_{acc}\]
     \begin{eqnarray}
      x_1(t)& = & \frac{1}{2}a t^2, \\
      x_2(t)& = & x_1(t) + L.
      \end{eqnarray}

     \[ t \ge t_{acc}\]
     \begin{eqnarray}
      x_1(t)& = & \frac{1}{2}a t_{acc}^2 + v(t - t_{acc}), \\
      x_2(t)& = & x_1(t) + L
      \end{eqnarray}
    where $v = a t_{acc}$.
 \par The space-time trajectories of O1 and O2 in S are shown in Fig.~4. At all times,
   the spatial separation of the objects is equal to their initial separation $L$.
   Because of the symmetry of the acceleration program, the separation of the objects 
   in the co-moving inertial frame S', $x'_2-x'_1$, is also $L$ at all times.
    \par The objects O1 and O2 may be replaced by clocks initially synchronised in
    the frame S, by the procedure described in the preceding section. In view of the
    symmetry of the acceleration program, it is evident that such clocks must remain
   synchronised at all times, including during the phase of uniform motion when
   $ t \ge t_{acc}$. The clocks are then mutually synchronised in the inertial frames S and S'
     at all times, but they are not necessarily synchronised with a clock at rest in S.
    \par A procedure is now established to synchronise a pair of clocks in S' with a
    pair of clocks in S at some instant in both S and S', whether or not the latter have been previously mutually synchronised
    by the procedure described in the previous section.  The first step towards this 
     goal is to set up an experiment to measure the distance between O1 and O2 during the 
     phase of uniform motion at speed $v$ of the two objects. As all such measurements
    in physics they are constructed from the coordinates corresponding to PMC. The method 
    employed is shown in Fig.~5. A simple ruler with two marks, $M1$ and $M2$ separated by the
    known distance $l$ is placed along the x-axis and the following PMCs are recorded
 using synchronised clocks in S at the positions of $M1$ and $M2$:
   \begin{eqnarray}
    PMC2(t_{21})  & \equiv & PO2(t_{21})@M1,~~~  PMC2(t_{22}) \equiv  PO2(t_{22})@M2, \nonumber \\
  PMC1(t_{11})  & \equiv & PO1(t_{11})@M1,~~~  PMC1(t_{12})  \equiv PO1(t_{12})@M2. 
    \end{eqnarray}
   Observation of the times of the four PMC above and knowledge of the ruler mark separation
   then enables both the spatial separation, $L$, of the objects and their speed, $v$, to be deduced
   from the equations derived from the space-time geometry of the four configurations,
   corresponding to the four $PMC$, shown in Fig.~5:
   \begin{eqnarray}
    t_{22}-t_{21} & = & \frac{l}{v}, \\
    t_{11}-t_{22} & = & \frac{L-l}{v}, \\
   t_{12}-t_{11} & = & \frac{l}{v}.  
    \end{eqnarray}
     The solution of these equations is:
   \begin{eqnarray}
    L & = & l \frac{(t_{12}+t_{11}- t_{22}-t_{21})}{(t_{12}-t_{11}+ t_{22}-t_{21})}, \\ 
     v & = & \frac{2  \ell}{(t_{12}-t_{11}+ t_{22}-t_{21})}. 
    \end{eqnarray}
  Now, in the case that $t_{11}= t_{22}$, Eq.~(4.9) gives:
    \begin{equation}
    L = l \frac{(t_{12}-t_{21})}{(t_{12}-t_{21})} =  l.
  \end{equation}
  i.e. the separation of the moving objects is equal to that of the ruler marks $M1$ and $M2$. 
   \par Inversely, if it is known in advance that the separation of the ruler marks is
     equal to that of the objects, observation of $PMC1(t_{11})$ and $PMC2(t_{22})$
   can be used to synchronise the clocks in S associated with the ruler marks $M1$ and $M2$. 
    More than this, if the objects O1 and O2 are now replaced by clocks in the frame S',
    the evident symmetry between the marks in S and pointers in S' and marks in S'
    and pointers in S, in this case,  means that the simultaneity of
     $PMC1(t_{11})$ and $PMC2(t_{22})$ {\it enables all four clocks, two in S, two in S' to be 
     mutually synchronised.} In practice this can conveniently be done by stopping all 
     four clocks and setting them to a common time. Denote the clocks, in an
    obvious notation, as C1(S),C2(S), C1(S') and C2(S'). The clocks  C1(S) and  C1(S')
    are started by  $PMC1(t_{11})$ and C2(S) and  C2(S') by  $PMC2(t_{22})$. Since $t_{11}= t_{22}$
     all four clocks are then mutally synchronised at a particular instant in both S 
     and S'. A possible practical way to implement
     this simultaneous starting of clocks in the frames S and S' using switches
     constructed from photon sources, photon detectors and screens is sketched in
     Fig.~6. 
\begin{figure}[htbp]
\begin{center}\hspace*{-0.5cm}\mbox{
\epsfysize15.0cm\epsffile{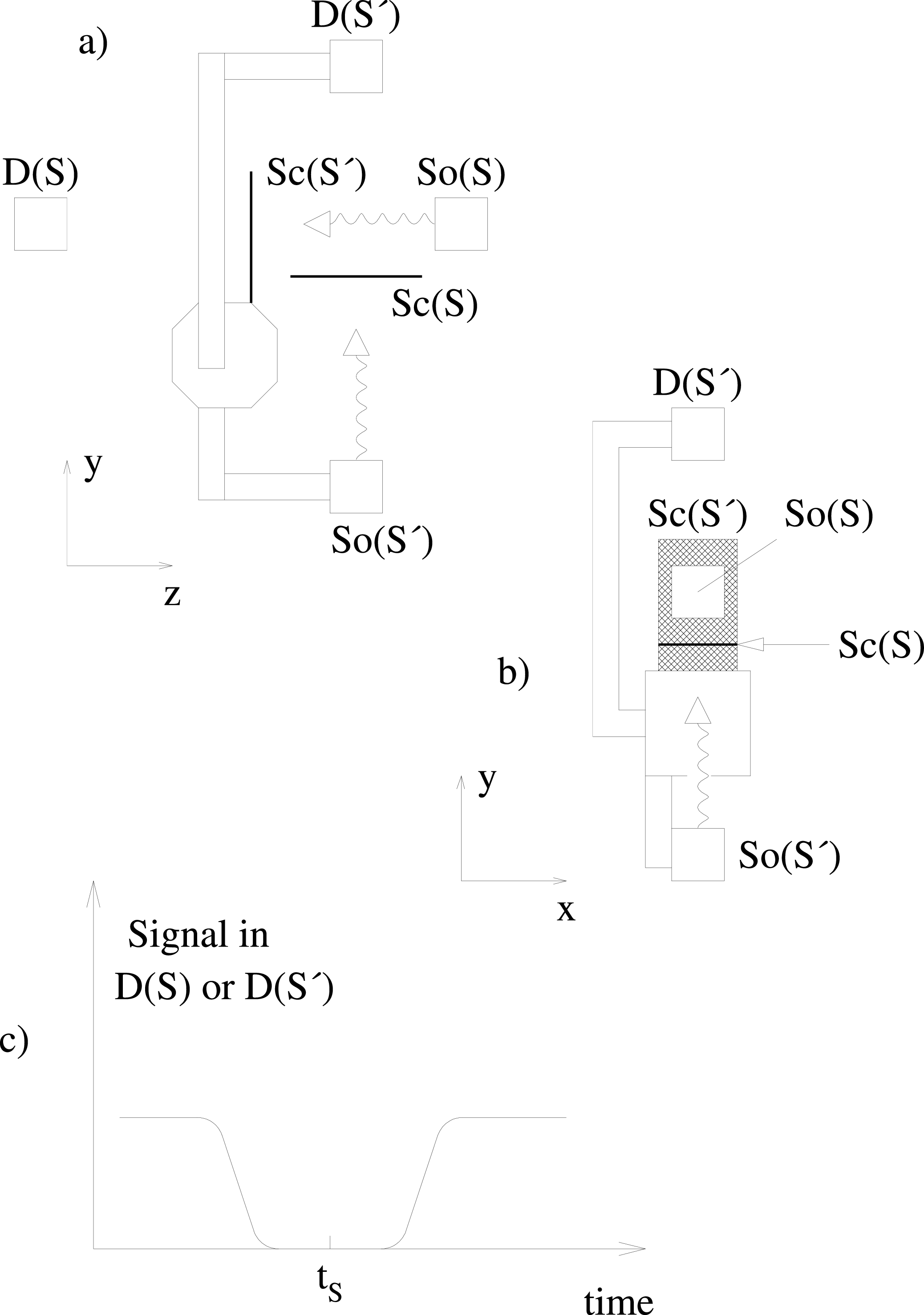}}
\caption{ {\em A practical scheme to synchronise clocks at rest in two inertial frames
  S and S' moving with relative velocity $v$ along their common x,x' axis.
   a) shows the z-y, b) the x-y projection
  of the apparatus. In both S and S' there are similar photon sources So(S) and So(S') and equidistant
  photon detectors D(S) and D(S'). When the clocks (not shown) connected directly to D(S) and D(S')
  are spatially separated the detectors D(S) and D(S') count continuously the photons emitted by
   So(S) and So(S'). Identically constructed screens Sc(S) and Sc(S') fixed in S and S'
   block both photon beams when source-detector pairs So(S)-D(S) and So(S')-D(S') (and hence
  the connected clocks) have the same x-coordinate, as shown in b). During the passage of the screens
  across the photon beams,
 the signal in D(S) or D(S') has the time variation shown in c). Clocks connected to D(S) and D(S'),
  with the same initial settings, when started at the time $t_S$, (which can be at any fixed position on the signal curve shown in c))
  are then synchronised.}}
\label{fig-fig6}
\end{center}
\end{figure}
     \par This method for synchronising clocks can also be used, instead of the procedure
          described in the previous section, to synchronise spatially separated clocks
           in the {\it same} inertial frame. This method, which may be called {\it clock
      synchronisation by length transport} can be graphically illustrated, as in Einstein's
       example demonstrating the importance of simultaneity in the definition of time
      measurements, by considering $PMC$ corresponding to train arrival times.
      On the platform at Cornavin station in Geneva there are two clocks, C1 and C2 
       that are unsynchronised but whose corresponding ruler marks $M1$ and $M2$ are a 
       known distance apart. A possible way to synchronise these clocks is as follows.
       They are both stopped and set to the same time, Apparata are set up to
     start them when suitable pointers are in spatial coincidence with their ruler
     marks. On the station at Bern there are two ruler marks the same distance
    apart as the clocks on the platform at Cornavin. Two identical locomotives
     are lined up so that corresponding points on them, $P1$ and $P2$, are aligned
     with these marks. The two locomotives then start at the same time and move along
     world lines of identical shape until they arrive at Cornavin station. When the
      points $P1$ and $P2$ of the locomotives are in spatial coincidence with the
      marks  $M1$ and $M2$:
      \[ PMC1 \equiv P1@M1,~~~ PMC2 \equiv P2@M2 \]
       the two clocks are started. They are synchronous. It is clear that any pair of 
       clocks along the line from Bern to Geneva, separated by the same distance as those
       at Cornavin, can be synchronised by the same technique.
\begin{figure}[htbp]
\begin{center}\hspace*{-0.5cm}\mbox{
\epsfysize15.0cm\epsffile{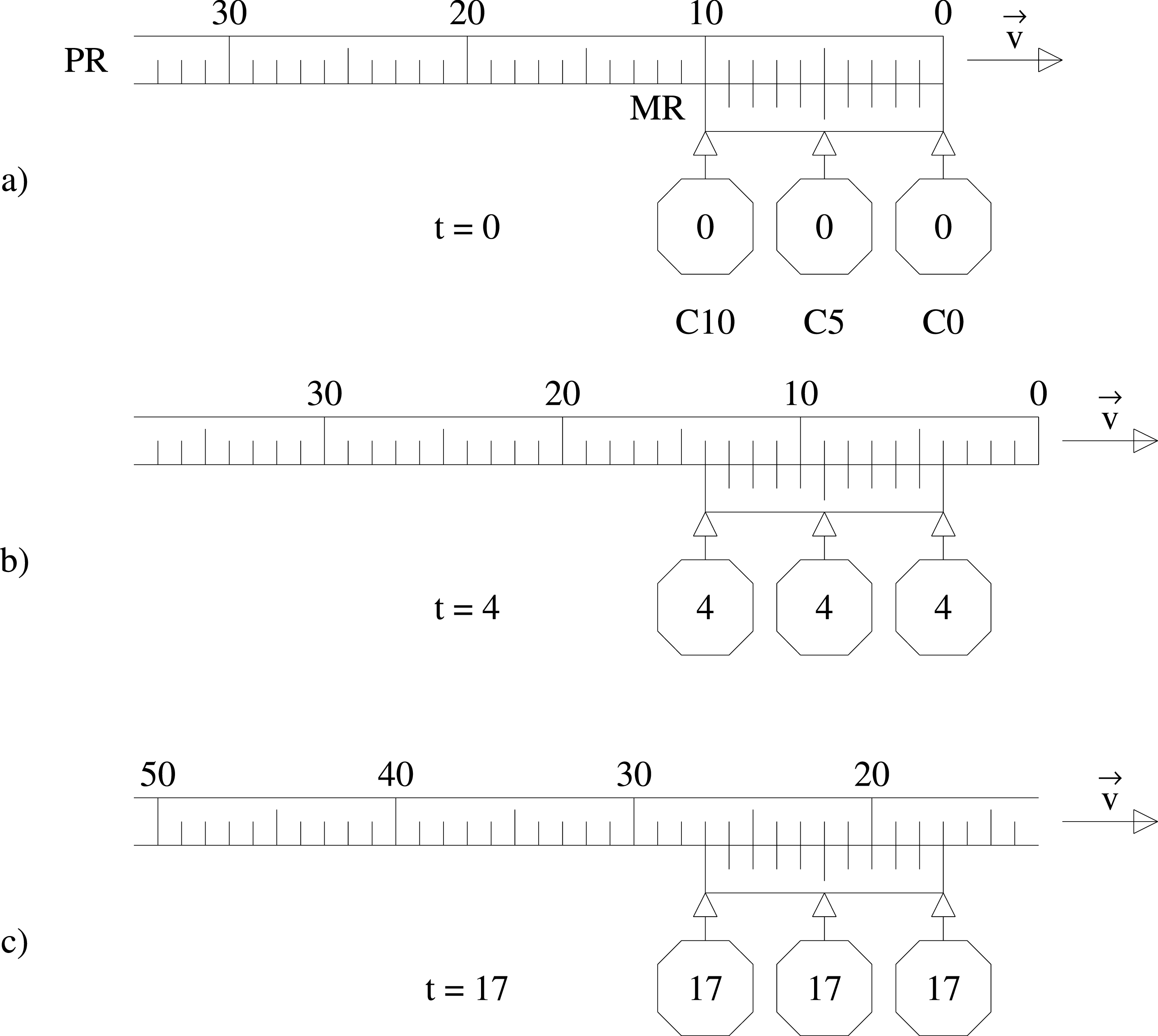}}
\caption{ {\em The moving ruler clock. The `pointer ruler' PR moves with uniform
  velocity $v$ parallel to the fixed `mark ruler' MR. The digital clocks
   C0, C5 and C10 which count the number of pointer coincidences with the marks
    $MR(0)$, $MR(5)$ and $MR(10)$ respectively constitute an array of spatially separated,
    synchronised, clocks in the rest frame of MR. a), b) and c) show the relative positions
    of PR and MR at different times in this frame.}}
\label{fig-fig7}
\end{center}
\end{figure}
       \par The method of clock synchronisation by length transport just described
        can be generalised to a large array of synchronised digital clocks in a given
        reference frame. The method is illustrated in Fig.~7. The `mark ruler' MR is at rest
        while the `pointer ruler' PR slides along MR at constant speed $v$. Opposite
       the mark $JM$ of MR is a detector of spatial coincidences of $JM$ with the pointer
       $JP$ constituted by the $JP$th mark of the moving ruler PR. Attached to the $PMC$ detector
        at $JM$ is the digital clock CJM. The configuration shown in Fig.~7a corresponds to the 
        following $PMC$, each with the structure $PMC(JM) \equiv PR(JP)@MR(JM)$:
        \[PMC(0) \equiv PR(0)@MR(0),~~~PMC(5) \equiv PR(5)@MR(5),\]
          \[PMC(10) \equiv PR(10)@MR(10). \]
          Each clock CJM contains a simple alogrithm to convert the corresponding pointer, $JP(JM)$
          of the ruler PR into a time:
         \begin{equation}
            t = \frac{d(JP(JM)-JM)}{v}
          \end{equation}
          where $d$ is the mark separation of MR or PR. The time $t$ given by (4.12) is independent
           of $JM$, that is, the whole array of clocks is synchronised. Further examples of the
           readings of the clocks C0, C5 and C10 are shown in Figs.7b and 7c.
           \par In practice, the $PMC$ detector does not need to identify $JP$ in order to
           construct the time, since the time interval recorded by each each clock 
           after the initial configuration of Fig.~7a is proportional to the number
           of $PMC$ recorded. Each clock is then simply a counter of the number of
           $PMC$s with a visual  display of its contents. These counters constitute
           an array of synchronised digital clocks with time resolution $\simeq d/v$.

\SECTION{\bf{The relation of measurements of space and time intervals
 in different inertial frames. The Galilean and Lorentz Transformations}}
      Suppose now that two clocks C1 and C2 are set beside the marks $M1$ and $M2$ in Fig.~5. 
 The objects O1 and O2 are replaced by clocks C1' and C2', at rest in S', an inertial
  frame moving with velocity $v$ along the positive x-axis in S, and separated by the
 same distance $L$ as C1 and C2 in the frame S. The positions of the clocks at rest in S', in S, 
at time $\tau$, are given, in general, by the relations\footnote{In the present
   and following sections the
     symbol $\tau$, $\tau'$ are used consistently to indicate the times of clocks
     at rest in S and S' as viewed by observers at rest in S and S', respectively,
     whereas $t$ and $t'$ are the times of clocks at rest in S and S' as 
     viewed by observers at rest in S' and S, respectively, The symbols $t$ and $t'$ therefore
     represent the apparent times registered by clocks in uniform motion.}:
  \begin{eqnarray}
  x(\rm{C1'}) - x_0 & = & v(\tau-\tau_0), \\
   x(\rm{C2'})& = & x(\rm{C1'})+L, \\
   x(\rm{C2}) & = & x(\rm{C1})+L.       
  \end{eqnarray}
 Similarly the positions of the clocks at rest in S, in S', at time $\tau'$, are:
  \begin{eqnarray}
  x'(\rm{C1}) - x'_0 & = & -v(\tau'-\tau'_0), \\
   x'(\rm{C2}) & = & x'(\rm{C1})+L, \\
   x'(\rm{C2'}) & = & x'(\rm{C1'})+L.       
  \end{eqnarray}
  The equations (5.1)-(5.6) do not assume any particular
  choice for the origins of spatial or  temporal coordinates. The proper times $\tau $ and $\tau'$
  are those recorded by synchronised clocks at any position in S and S' respectively.
   In addition to these `frame times', the following
   `apparent' times may also be defined:
   \par $t'({\rm C1'},\tau)$: The time registered by C1' seen by a local observer in S at time $\tau$.
   \par $t'({\rm C2'},\tau)$: The time registered by C2' seen by a local observer in S at time $\tau$.
   \par $t( {\rm C1},\tau')$: The time registered by C1 seen by a local observer in S' at time $\tau'$
   \par $t({\rm C2},\tau')$: The time registered by C2 seen by a local observer in S' at time $\tau'$
    \par It is important to point out that (5.1)-(5.6) do not represent events that are connected by a
     space-time transformation but rather specify kinematical configurations in the frame S of
     a {\it primary experiment} and in the frame S' of a physically independent
      {\it reciprocal experiment}. Thus different space-time transformations describe the connection
       between events in the frames S and S', in the primary and reciprocal experiments\footnote{An example
      of this is given below where the LT (5.17) and (5.18) applies for the primary
          experiment and the different LT (5.24) and (5.25) applies for the reciprocal one.}
       Hence the time $t'$ is related to $\tau$
       via a certain space-time LT in the primary experiment and $t$ to $\tau'$ by a {\it different} 
         space-time LT in 
      the reciprocal experiment.   
       The configurations: S' moves with speed $v$ along the positive $x$-axis in S (primary experiment) and
        S moves with speed $v$ along the negative $x'$-axis in S' (reciprocal experiment) are, however,
      related by the {\it kinematical}
       (velocity) LT between these two frames.
   \par The relations (5.1)-(5.6)  may be simplified by a suitable choice of coordinate origins
    and synchronisation of the frame clocks in S and S'. All four clocks are now synchronised
    using the length transport procedure described in the previous section. The corresponding
    pointer-mark coincidences (for simplicity of notation the clock labels are used to
    denote either pointers or marks) are as follows:
     \[ PMC1'(\tau)\equiv C1'(\tau)@C1,~~~~PMC2'(\tau)\equiv C2'(\tau)@C2, \]
      \[ PMC1(\tau')\equiv C1(\tau')@C1',~~~~PMC2(\tau')\equiv C2(\tau')@C2'. \]
   As shown in Fig.~8a, the choice $\tau= \tau' = 0$ enables synchronisation of all four clocks at these
   times.
   With the choice of coordinate origins shown in Fig.~8
a, (5.1)-(5.6) simplify to:
  \[ primary~experiment \]
  \begin{eqnarray}
  x(\rm{C1'}) & = & v \tau, \\
   x(\rm{C2'}) & = & x(\rm{C1'})+L, \\
   x(\rm{C1}) & = & 0, \\
   x(\rm{C2}) & = & L.       
  \end{eqnarray}
 \[ reciprocal~experiment \]
  \begin{eqnarray}
  x'(\rm{C1}) & = & -v \tau', \\
   x'(\rm{C2}) & = & x'(\rm{C1})+L, \\
 x'(\rm{C1'}) & = & 0,  \\
   x'(\rm{C2'}) & = & L.       
  \end{eqnarray}
  As in (5.1)-(5.6) above it is assumed that all clocks are observed to run at the same rate
   when they are in the same inertial frame.
   The observed behaviour of the moving clocks in the primary and reciprocal 
   experiments is specified by the four apparent times
   $t'(\rm{C1'},\tau)$, $t'(\rm{C2'},\tau)$, $t(\rm{C1},\tau')$ and $t(\rm{C2},\tau')$
    defined above. In the case that all these times are equal to $\tau$, which is in turn
    equal to $\tau'$, for all values of $\tau$, time is universal, for all inertial
    observers, as Newton assumed.  The GT formulae for the primary experiment are more conventionally
    written for a single (`stationary') clock in S recording time $\tau$ and a (uniformly moving, as viewed
     from S) one, at the origin in S', recording time $t'$,
    as:
    \[ primary~experiment \]
   \begin{eqnarray}
      x' & = & x - v \tau = 0, \\
      t' & = & \tau.
   \end{eqnarray}
    The LT of the times and positions of single clocks in S and S'
    in the primary experiment is written in a similar manner to the GT (5.15), (5.16) as (equations (A.44) and (A.45)
     in the Appendix):
  \[ primary~experiment \]
   \begin{eqnarray}
      x' & = & \gamma(x - v \tau) = 0,  \\
    t'(\tau) & = &\gamma(\tau-\frac{vx}{V^2})
    \end{eqnarray}
     where 
      \[  \gamma \equiv \frac{1}{\sqrt{1-(\frac{v}{V})^2}}. \]
     and $V$ is the maximum possible relative velocity of the frames S and S'.
    This particular choice of coordinate origins and time offsets, introduced
    by Einstein, corresponds to synchronisation of the clocks in S and S'
    so that $\tau= t'(\tau) = 0$ when $x = x' =0$. It reduces to the GT
    (5.15) and (5.16) in the limit $V \rightarrow \infty$. LT equations suitable to
     describe the clocks synchronised as shown in  Fig.~8a may be derived by
    generalising (5.17) and (5.18) to allow an arbitary choice of coordinate origins
    and clock-time offsets:
 \[ primary~experiment \]
   \begin{eqnarray}
      x'- x'_0 & = & \gamma(x-x_0 - v (\tau-\tau_0)) = 0 , \\
      t'(\tau)-t'_0 & = & \gamma(\tau-\tau_0-\frac{v(x-x_0)}{V^2}).
    \end{eqnarray}
     The importance of such a generalisation, in order to describe synchronised
      clocks at different spatial positions, was clearly stated by Einstein in
      the original paper on SR~\cite{Ein1}:
  \par {\tt If no assumption whatever be made as to the initial position of the moving system
   and as the the zero point of $\tau$,} ($t'$ in the notation used above){\tt  an additive constant
   is to be placed on the right side of each of these}~(the LT (5.17)-(5.18)) \newline
{\tt equations.}
   \par This consideration was never, however, to the present writer's best knowledge, 
    taken into account by Einstein, or any
     later author, before the work presented in Ref.~\cite{JHF3}. The `standard' LT
      (5.17)-(5.18) has always been employed, regardless of the the position
      of the clock in the frame S'. How this results in the spurious predictions of
      `relativity of simultaneity' and `length contraction' effects is explained in the following
      section.
\begin{figure}[htbp]
\begin{center}\hspace*{-0.5cm}\mbox{
\epsfysize15.0cm\epsffile{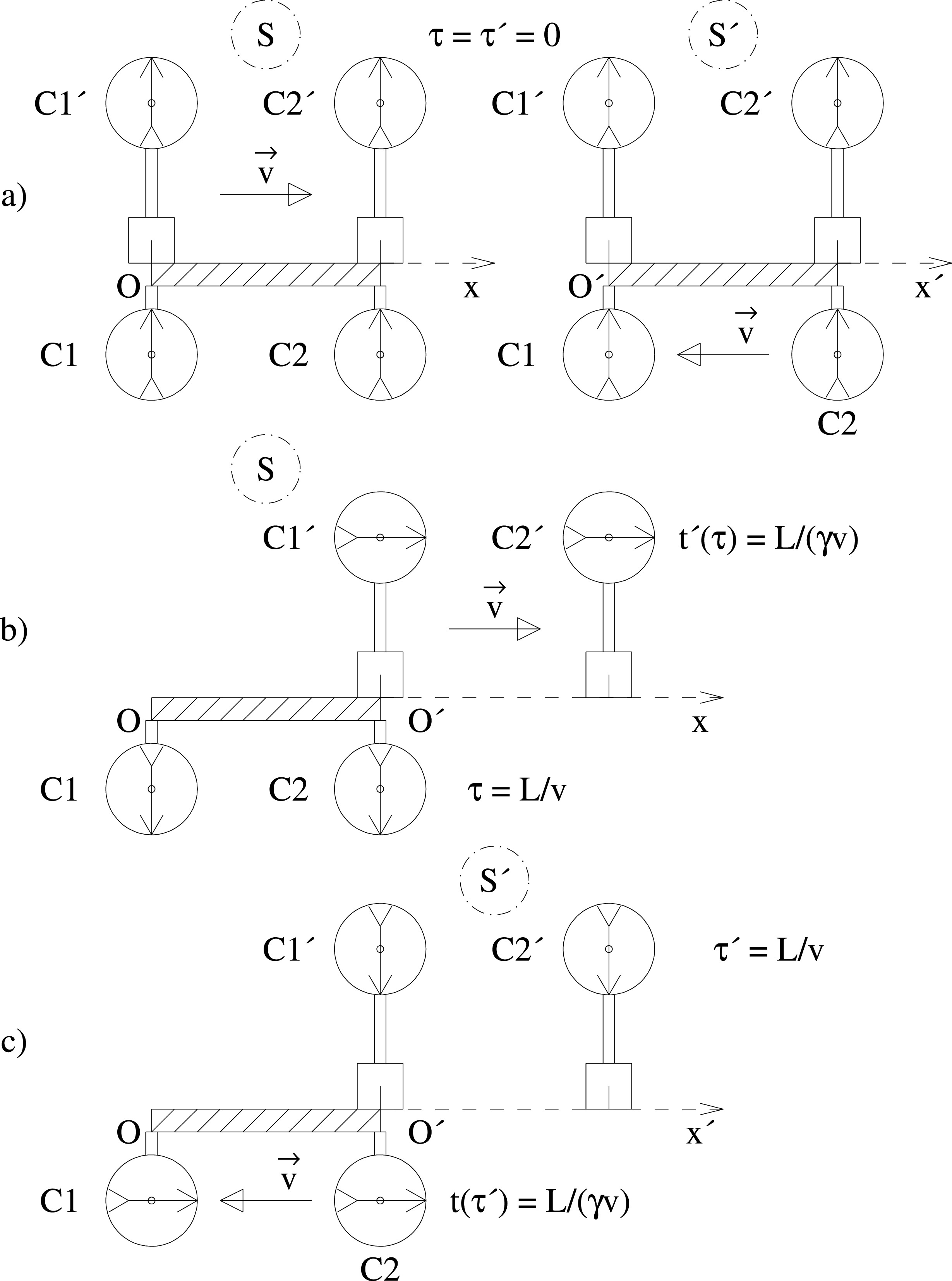}}
\caption{ {\em  The clocks C1 and C2 are separated by the distance $L$ in S, C1' and C2'
   by the same distance in S'. a) All four clocks are synchronised by length transport (see text)
  at the times $\tau = \tau' =0$ either in the the frame S (primary experiment)
     or in the frame S' (reciprocal experiment). b) Times seen by an observer at rest in S
    in the primary experiment at frame time $\tau = L/v$.
    c) Times seen by an observer at rest in S' in the reciprocal experiment at frame time $\tau' = L/v$. $\beta = v/V = \sqrt{3}/2$,
     $\gamma = 2$. Units of distance and time are chosen so that $L/V = 3\sqrt{3}$.}}
\label{fig-fig8}
\end{center}
\end{figure}

     Since (see Fig.~8a), $\tau({\rm C1})= t'({\rm C1'}) = 0$ when $x({\rm C1}) =  x'({\rm C1'})= 0$
  Eq.~s(5.17) and (5.18) already correctly describe the synchronisation of C1 and C1'.
   i.e. in (5.19) and (5.20) the choice $x_0 = x'_0 = t_0 =  t'_0 = 0$ must be made.
   Substituting the relation:
  \[ x'({\rm C1'}) = 0 = x({\rm C1'})-v \tau \]
   into(5.18) gives:
    \[t'({\rm C1'},\tau) = \gamma \left( \tau -(\frac{v}{V})^2 \tau \right) = \frac{\tau}{\gamma}. \]
     The transformation equations for the clock C1' in the primary experiment are then:
 \[ primary~experiment \]
  \begin{eqnarray}
  x({\rm C1'}) & = & v \tau, \\
 x'({\rm C1'}) & = & 0,  \\
 t'({\rm C1'},\tau) & = & \frac{\tau }{\gamma}.
  \end{eqnarray}
    In a similar manner, substituting the relation:
  \[ x({\rm C1}) = 0 = x'({\rm C1})+v \tau' \]
 into the equations, reciprocal\footnote{Note that (5.24) and (5.25) are reciprocal to, not
    the inverses, of (5.17) and (5.18), since in (5.17) and (5.18) $x'= 0$, $x = v \tau$, 
    while in (5.24) and (5.25) $x= 0$, $x' = -v \tau'$. S and S' are therefore related by  
      different space-time LTs in the primary and reciprocal experiments.} to (5.17) and (5.18):
 \[ reciprocal~experiment \]
     \begin{eqnarray}
      x & = & \gamma(x' + v \tau') = 0,  \\
      t(\tau') & = & \gamma(\tau'+\frac{v x'}{V^2})
    \end{eqnarray}
 gives:
    \[t({\rm C1},\tau') = \gamma \left( \tau' -(\frac{v}{V})^2 \tau' \right) = \frac{\tau'}{\gamma} \]
  thus yelding for the transformation equations of the clock C1 in the reciprocal experiment:
 \[ reciprocal~experiment \]
       \begin{eqnarray}
 x'({\rm C1}) & = & -v \tau', \\
 x({\rm C1}) & = & 0, \\
 t({\rm C1},\tau') & = & \frac{\tau'}{\gamma}.
    \end{eqnarray}
  Comparing with Eq.~s(5.7) and (5.11)  it can be seen
  that the space transformations are the same as for the GT 
   whereas the apparent times $t'(\rm{C1'},\tau)$ and $t(\rm{C1},\tau')$
   differ from $\tau$ and $\tau'$ respectively. Observers in S in the primary experiment and in S'
   in the reciprocal experiment judge
   that the clock in the other frame is running slow by the factor $1/\gamma$ ---the time
   dilation effect first derived as a general consequence of the LT, by Einstein~\cite{Ein1}.
  
   \par In order to describe correctly the synchronisation of C2 and C2', a different
    choice of constants in (5.19) and (5.20) is necessary. Suppose that  $\tau=  t' = 0$
    when $x = L$ and $x'= L'$. This implies that $\tau_0 =  t'_0 = 0$ and $x_0 = L$, $x'_0 = L'$
    so that:
 \[ primary~experiment \]
  \begin{equation}
  x'- L' = \gamma(x-L - v \tau) = 0,
   \end{equation}
  \begin{equation}
  t'(\tau) = \gamma( \tau -\frac{v(x-L)}{V^2}).
   \end{equation}
   Since both $x_0$ and  $x'_0$ are constants, independent of $v$, and depending only on the
   choice of spatial coordinate systems in S and S' respectively, Eq.~(5.29) holds for all values
   of $v$; in particular it holds when $v \rightarrow 0$, S $\rightarrow$ S', $x \rightarrow x'$
   in which case (5.29) gives:
  \begin{equation}
 x'-L' = x' -L
  \end{equation}
 or
  \begin{equation}
     L' = L.
  \end{equation}
 As previously discussed, from first principles, in Section 4 above the spatial separation
 of the clocks is a Lorentz invariant quantity ---there is no `relativistic length contraction'.
 With (5.32) and defining $X \equiv x-L$ and  $X' \equiv x'-L$ (5.29) and (5.30) 
   are identical to (5.17) and (5.18) with the replacements $x \rightarrow X$ and 
   $x'\rightarrow X'$. The transformation equations for C2' and C2 are then:
\[ primary~experiment \]
  \begin{eqnarray}
  x(\rm{C2'}) & = & v \tau + L=  x(\rm{C1'}) + L, \\
 x'(\rm{C2'}) & = & L= x'(\rm{C1'}) +  L, \\
 t'(\rm{C2'},\tau)& = & \frac{\tau}{\gamma}= t'(\rm{C1'},\tau).
 \end{eqnarray}
 A similar calculation for the reciprocal experiment gives:
\[ reciprocal~experiment \]
  \begin{eqnarray}
 x'(\rm{C2}) & = & -v \tau'+ L = x'(\rm{C1}) + L, \\
 x(\rm{C2}) & = &L  =   x(\rm{C1}) + L, \\
 t(\rm{C2},\tau') & = & \frac{\tau'}{\gamma} =  t(\rm{C1},\tau').
 \end{eqnarray}
  Eqs.~(5.33),(5.34) and (5.36),(5.37) are the same as the corresponding GT 
  whereas the apparent times of C1' and C2' as viewed from S in the primary experiment are equal,
  as are the apparent times of C1 and C2 as viewed from S' in the reciprocal experiment.
   \par The value of the coordinate offset, $x_0 = L$, necessary to ensure synchronisation 
    of C2 and C2', is equivalent to choosing a LT which is `local at C2', in the sense that
    C2' is placed at the origin of spatial coordinates in S'. This is so because $X' = x'- L$ actually
    vanishes when $x'$ specifies the position of C2'. Clearly all such clocks, (i.e., whatever
    the value of $L$) described by a local LT or, equivalently, by Eq.~s(5.29) and (5.30), will be
    synchronous with
    C1' at $\tau = \tau'= 0$ (the instant of synchronisation), as well as at all later times.
     The necessity to use, in a systematic manner, such `local' LT to describe clocks in order
    to avoid causal paradoxes, and to assure unique predictions in special relativity that
     respect translational invariance, has been previously pointed out and discussed in detail in
     Reference~\cite{JHF3}. 
   \par The observed behaviour of the clocks C1', (C1) as predicted
     by Eq.~s(5.21)-(5.23), (Eq.~s(5.26)-(5.28)) or of C2', (C2)
      as predicted by Eq.~s(5.33)-(5.35), (Eq.~s(5.36)-(5.38))
     is illustrated in Fig.~ 8.
    It is assumed, for purposes of illustration, that $\beta \equiv v/V = \sqrt{3}/2$, $\gamma = 2$,
     so that the apparent rate of the moving clocks is $50 \%$  less than that of a similar
    clock viewed at rest. Units of length and time are chosen such that $L = 3\sqrt{3}V$. Twelve  
    units of time then correspond to a $2 \pi$ rotation of the hands of the analogue clocks in Fig.~8. 
    As in Fig.~1d, there are twelve marks on the faces of the clocks. Fig.~8a shows the clocks 
     as viewed from either S in the primary experiment or S' in the reciprocal
     experiment at the instant, $\tau= \tau'= 0$, at which they are synchronised.
      At time $\tau= L/v$ the
    following five $PMC$ may be recorded by observers in S (Fig.~8b):
\[ primary~experiment \]
    \begin{itemize} 
     \item Spatial coincidence of C1' (pointer) with C2 (mark):
             \[ PMC1 \equiv C1'(L/v) @ C2.\]
             
    \item Pointer-mark coincidence of clock C1:
           
            \[  PMC2 \equiv PC1(L/v) @ MC1(6). \]
              
    \item Pointer-mark coincidence of clock C2:
           
           \[   PMC3 \equiv PC2(L/v) @ MC2(6). \]

   \item Pointer-mark coincidence of clock C1':
          
          \[    PMC4 \equiv PC1'(L/v) @ MC1'(3). \]
              
  \item Pointer-mark coincidence of clock C2':
         
         \[     PMC5 \equiv PC2'(L/v) @ MC2'(3). \]
            
   \end{itemize}
    The time $\tau$ of C1 and C2 in S in the primary experiment is given by the marks $MC1(6)$ or $MC2(6)$
    while the apparent  time of C1' or C2' as viewed from S, $t'(\tau)$, is given by
    $MC1'(3)$ or $MC2'(3)$. That is, with the given choice of space and time units:
           \begin{equation}
         \tau= 6,~~~t'(\tau) = t'(6) = 3.
      \end{equation}
    The situation at the instant of the same spatial coincidence of C1' and C2, but as viewed from
    S' in the reciprocal experiment, is shown in Fig.~8c. Again, observations of five distinct $PMC$ may be made. They are:
\[ reciprocal~experiment \]
    \begin{itemize}
 
     \item  Spatial coincidence of C2 (pointer) with C1' (mark):
            
            \[  PMC1' \equiv C2(L/v) @ C1'. \]
    \item Pointer-mark coincidence of clock C1':
          
          \[    PMC2' \equiv PC1'(L/v) @ MC1'(6). \]
            
    \item Pointer-mark coincidence of clock C2':
            
           \[   PMC3' \equiv PC2'(L/v) @ MC2'(6). \]

   \item Pointer-mark coincidence of clock C1:
           
           \[   PMC4' \equiv PC1(L/v) @ MC1(3). \]
             
  \item  Pointer-mark coincidence of clock C2:
          
        \[    PMC5' \equiv PC2(L/v) @ MC2(3). \]
    \end{itemize}            
  The times in S' corresponding to the times in S given by (5.39) are:
   \begin{equation}
         \tau' = 6,~~~t(\tau') = t(6) = 3.
   \end{equation}
  The times shown in (5.39) and (5.40) are a perfect exemplification
  of the measurement reciprocity postulate stated in the Introduction:
   space-time measurements in reciprocal experiments produce identical results. Observers in
     both the primary and reciprocal experiments see that the 
    moving clocks in the other frame are running more slowly than similar clocks 
    at rest in their own frames.

 \SECTION{\bf{The illusory nature of `relativity of simultaneity' and `relativistic
   length contraction' of conventional special relativity theory}} 

   In the length transport synchronisation of the clocks C1, C2 , C1' and C2' in Fig.~8a
   it was noticed that different values of the constants $x_0$, $x'_0$, $\tau_0$ and  $t'_0$
    in the general LT equations (5.19) and (5.20) were needed to correctly describe the
    positions and times of the spatially coincident clock pairs C1,C1' and C2,C2'. For
    C1,C1', $x_0 = x'_0 = \tau_0 = t'_0 =0$, while for  C2,C2' $x_0 = x'_0 = L$,  $\tau_0 = t'_0 = 0$. 
    If it is now attempted to describe the times of the clock pair  C2,C2' with the LT appropriate
    to  C1,C1' i.e. by using the standard LT of Eq.~s  (5.17)  and (5.18) in the 
     primary experiment and  Eq.~s (5.24)  and (5.25) in the reciprocal experiment, denoted as LT(0), the frame times
     $\tau$ and $\tau'$ and the apparent times $t'(\tau)$ and $t(\tau')$ are those shown in the
      first and second 
     rows of Table 1 (for C1 and C1') and  Table 2 (for C2 and C2')\footnote{Note that in (5.17),        
   $x'= 0$ and in (5.24) $x= 0$. When the coordinates of C2 or C2'are substituted in these
   equations, these values are replaced by  $x'= L$ and  $x= L$ respectively. That is, a formal
   coordinate substitution is made in the first members of (5.17) and (5.24) without regard 
   to the physical meaning of these equations, that describe correctly only the clocks C1' and C1
   respectively.}.
       The first row in each table shows the apparent times $t'(\tau)$ of the clocks
     at rest in S' as viewed from S in the primary experiment, whereas the second row contains the apparent times
      $t(\tau')$ of the clocks at rest in S as viewed from S' in the reciprocal experiment.
       Also given in Tables 1 and 2 are
     the positions of the `moving' clocks in either `stationary' frame as derived by applying LT(0),
    or its inverse, to all the clocks.

\begin{figure}[htbp]
\begin{center}\hspace*{-0.5cm}\mbox{
\epsfysize12.0cm\epsffile{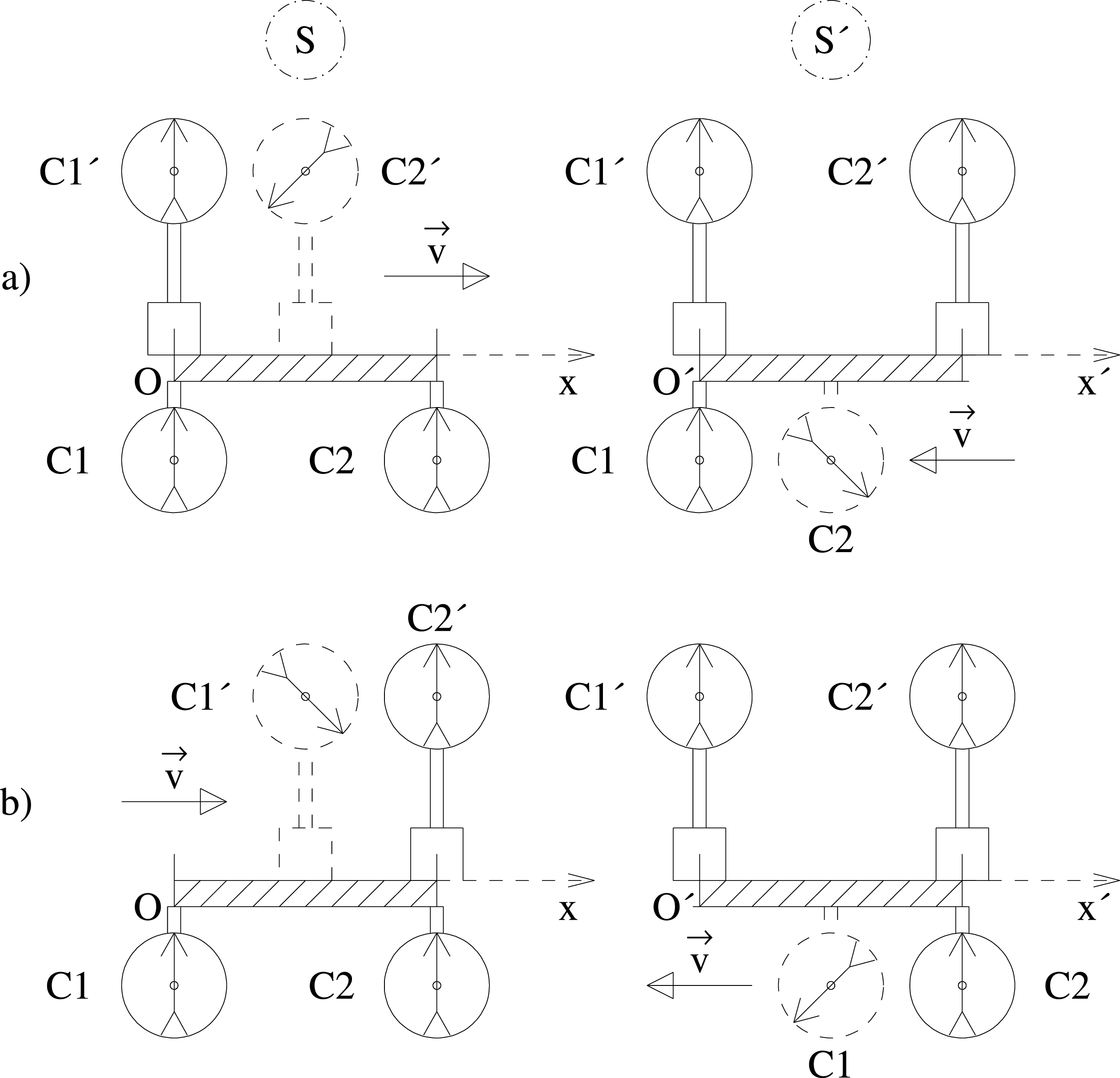}}
\caption{ {\em Examples of misuse of the space-time LT leading to the spurious `relativity of simultaneity'
   and `length contraction' effects in primary experiment (frame S) and in
     the reciprocal experiment (frame S'). The clocks C1, C2, C1' and C2 are assumed to be initially synchronised
   by length transport as shown in Fig.~8a. In a), LT(0), appropriate for C1 and  C1' is applied also to
   C2 and C2'. In S, C1 and C2 are assumed to be synchronised at $t=0$. Setting $x' = L$ for
   C2' in LT(0) then gives the apparent position (dashed) and time for C2' shown. In S' C1' and C2'
    are assumed to be synchronised at $\tau'=0$.  Setting $x = L$ for C2 gives the apparent position
     (dashed) and time for C2 as shown. The distance between C1' and the apparent position of
    C2' in S shows the `length contraction' effect. Also the apparent time of C2' is not
     synchronous with the frame times of C1 and C2. This is the `relativity of simultaneity' effect.
    Reciprocal effects  are seen in a) by the observer in S'. In b) the transformation LT(L), appropriate
   for C2 and C2' is
   applied also to C1 and C1'. As in a), `length contraction' occurs between the pairs C1',C2' viewed
  from S and  C1,C2 as viewed from S', but now it is C1,C1' instead of C2,C2', as in a) that show
   `relativity of simultaneity'. See the text for further discussion.
    Parameter values and units as in Fig.~8.}}
\label{fig-fig9}
\end{center}
\end{figure}
\begin{table}
\begin{center}
\begin{tabular}{|c|c||c c c c|c c c c|} \cline{3-10} 
\multicolumn{2}{c}{ }& \multicolumn{4}{|c|}{C1} &  \multicolumn{4}{c|}{ C1'} \\ \hline
   Obs & LT & $x$ & $x'(\tau')$ & $\tau$ & $t(\tau')$ &  $x'$ & $x(t)$ & $\tau'$ & $t'(\tau)$ \\ \hline \hline
  S & LT(0) & --  & -- & 0 & -- & 0 & 0 & -- & 0 \\ \hline
  S' & LT(0) & 0 & 0 & -- & 0 & -- & -- & 0 & -- \\ \hline
  S & LT(L) & -- & -- & 0 & -- & 0 & $L(\gamma -1)/\gamma$ & -- & $v L/V^2$ \\ \hline
  S' & LT(L) & 0 & $L(\gamma -1)/\gamma$ & -- & -$v L/V^2$ & -- & -- & 0 & -- \\ \hline
\end{tabular} 
\caption[]{{\em  Apparent times and positions of the clocks C1,C1' in Fig.~9.

   LT(0) corresponds to the LT of Eqs.~(5.17) and (5.18) (primary experiment) or
      (5.24) and (5.25) (reciprocal experiment). LT(L) corresponds to
      the same equations with the replacements: $x \rightarrow x-L$,  $x' \rightarrow x'-L$.
     For an observer in S in
    the primary experiment the values
  of $\tau$ and $x'$ are assigned the values shown while $x(t)$  and $t'(\tau)$ are calculated using the LT.
   For a observer in S' in the reciprocal experiment the values  of $\tau'$ and $x$  are assigned  the values shown while 
   $x'(t')$  and $t(\tau')$ are calculated using the LT. See text for discussion.}}     
\end{center}
\end{table}

\begin{table}
\begin{center}
\begin{tabular}{|c|c||c c c c|c c c c|} \cline{3-10} 
\multicolumn{2}{c}{ }& \multicolumn{4}{|c|}{C2} &  \multicolumn{4}{c|}{ C2'} \\ \hline
   Obs & LT & $x$ & $x'(t')$ & $\tau$ & $t(\tau')$ &  $x'$ & $x(t)$ & $\tau'$ & $t'(\tau)$ \\ \hline \hline
  S & LT(0) & -- & -- & 0 & -- &  $L$ &  $L/\gamma$ & -- & -$v L/V^2$ \\ \hline
  S' & LT(0) & $L$ & $L/\gamma$  & -- & $v L/V^2$ &  -- & -- & 0 & -- \\ \hline
  S & LT(L) &  -- & -- & 0 & -- &  $L$ & $L$ & -- & 0 \\ \hline
  S' & LT(L) & $L$ &  $L$ & -- & 0  & -- & -- & 0 & -- \\ \hline
\end{tabular} 
\caption[]{{\em  Apparent times and positions of the clocks C2, C2' in Fig.~9. calculated
     as described in the Caption of Table 1. See text for discussion.}}     
\end{center}
\end{table}

     \par For example, application of LT(0) to C1 and C1' at $t = 0$ gives $ x = x' = 0$ and $t'(t) = 0$.
       Setting $\tau = 0$ and $x'= L$ in the inverse of (5.18):
      \begin{equation}
           \tau = \gamma(t'(\tau)+\frac{v x'}{V^2})
      \end{equation}
       (i.e. now applying LT(0) to the clock C2') gives $t'(\tau) = -vL/V^2$ Substituting this value for
       $\tau'$ and $x' = L$ into the the inverse of (5.17):
      \begin{equation}
       x = \gamma(x'+v \tau')
      \end{equation}
      gives the `apparent' position  $x(t)$ of C2' of
       \begin{equation}
        x(0) = \gamma L(1-\frac{v^2}{V^2}) = \frac{L}{\gamma}.
      \end{equation}  
       The difference between the position $x = 0$ of C1 or C1' and the above apparent position
       of C2' is then $L/\gamma$. This is the well-known relativistic `length contraction' effect.
        It is seen to be a spurious consequence of the application of LT(0) appropriate for the
      synchronised clocks C1 and C1' also to the clocks C2 and C2' for which 
      the synchronous behaviour shown in Fig.~8a requires the LT (5.29) and (5.30) in 
       the primary experiment and corresponding equations in the reciprocal experiment, denoted as LT(L).
      The entries in the second rows of Tables 1 and 2 are obtained by inverting the roles of the frames
       S and S' in the calculation just performed, that is, for the primary experiment,
       by substituting $\tau' = 0$ and $x = L$ into 
       LT(0). The entries in the third and fourth rows of Table 1 are the results of performing
      analogous calculations where the correct transformation for C2 and C2', LT(L), is also
      applied (incorrectly) to C1 and C1'.
        The entries of Table 1 and 2 are displayed visually in Fig.~9 where the spurious
      `apparent' positions of the moving clocks are indicated by dashed lines.
      \par The so-called `relativity of simultaneity' effect of Einstein's special relativity theory
      is apparent in the entries of Table 1 and 2 and in Fig.~9. In Fig.~9a the observer in S in the
        primary experiment sees that
   the clocks C1 and C2, at rest in his own frame, are synchronous, but that that the moving clocks
         C1' and C2' are not. In contrast the observer in S' in the reciprocal experiment
     judges C1' and C2' to be synchronous,
         but not C1 and C2 that are in motion relative to him. Thus events that are synchronous 
      in one frame (say, particular times of C1 and C2 as viewed by the observer at rest in S) are not so 
     in another frame (any times indicated by the same two clocks when viewed by an observer at rest in 
       S'). 
       Since an identical  and simultaneous synchronisation procedure
       is applied
       to the clock pairs  C1,C1' and C2,C2' the behaviour shown in Fig.~9a is clearly 
       wrong; it is simply the consequence of using constants in the LT equations for
       C2 and C2' that do not correctly describe the synchronisation of all four 
       clocks as shown in Fig8a. It is clear from inspection of Eqs(5.23),(5.33) and (5.28) (5.36)
       that when the constants in the LT are chosen so as to correctly describe the
       synchronisation procedure, the time ordering of events is the same in any
      inertial frame --there is, in this case, no `relativity of simultaneity'. Indeed, this is necessary
      if causal paradoxes\footnote{Some examples are discussed in Reference~\cite{JHF4}.}
       are to be avoided.
      \par In Fig.~9b, are shown the results obtained when the correct LT for the clocks C2, C2', LT(L), is
       also (incorrectly) applied to
       the clocks C1 and C1'. In this case C1' is not synchronised with the clocks in S in the
       primary experiment at $\tau = 0$,
       and C1 is not synchronised with the clocks in S' in the reciprocal experiment
       at $\tau' = 0$. Again, the imposed synchronisation
       of all four clocks in both S and S' as shown in Fig.~8a is not respected.
       \par Comparing the apparent times of C1' and C2', as viewed from S, in Fig.~9a and Fig.~9b, it can
       be seen that C1' is in both cases in advance of C2' by 4.5 time units. However, in Fig.~9a, C1'
       shows the same time as C1 and C2, whereas in Fig.~9b it is C2' that shows the same time as
       C1 and C2. Also, although the apparent distance between C1' and C2' is always $1/\gamma$ times
      the distance between C1 and C2, the positions of C1' and C2' relative to C1 and C2 are different in
      the two cases. In Fig.~9a, C1 and C1' are aligned and C2' is shifted from C2, whereas in Fig.~9b,
       C2 and C2' are aligned and C1' is shifted from C1. So, although the size and sign of the `relativity
       of simultaneity' effect and the size of the `length contraction' effect are the same in Fig.~9a
       and Fig.~9b, the observer in S sees that the clocks C1' and C2' are in different positions, and
       show different times, in the two cases. Applying instead the Lorentz transformation LT(L/2) to
        C1' and C2' in Fig.~9a instead of LT(0) results again in the same  `relativity
       of simultaneity'  and `length contraction' effects, but the apparent positions of C1' and C2'
       are found to be equidistant from C1 and C2 respectively, and neither  C1' nor C2'  is seen to
       be synchronous
       with C1 and C2. In fact, as shown in Reference~\cite{JHF3}, by a suitable choice of the value
       of the parameter $\alpha$ in LT($\alpha$L) given by Eq.~s(5.29) and (5.30) ---i.e. a different
       choice of coordinate orgin in S ---  the apparent position of, say, 
       C2' in Fig.~9a, can be situated at ~{\it any position} along the x-axis! Since evidently the observed
       positions of the clocks C1' and C2' cannot depend, in this way, on an arbitary choice of the
       coordinate origin in S, all the configurations shown in Fig.~9a as well as those just discussed
       are spurious and unphysical. The correct physical situation, according to the LT and
       the defined synchronisation procedure at $t = 0$ is that shown in Fig.~8a. 
       There are then no `relativity of simultaneity' and `length contraction' effects in this example.
 
 \SECTION{\bf{Light at last}} 
  The discussion of the previous sections concerned rigorous definitions
   of the measurement of space and time intervals and the ways in which the results of
  such measurements are related via Galilean or Lorentz transformations. There is no overlap,
  after the important remark, quoted above, concerning the measurement of time and the concept
   of simultaneity, with the approach of Einstein in his first special relativity paper.
   Einstein postulated that the speed of light has the same value in any inertial frame
   and that it does not depend on the motion of its source. This highly counter-intuitive
   property of light is the crux of the conceptual difficulty of special relativity
 --the point at which classical `commonsense' breaks down. It may be compared, in this respect,
  with the superposition principle of quantum mechanics. The reader of the present paper
  has understood the valid conclusion of Einstein's paper concerning space-time geometry:
  the time dilation effect, and also why some other predictions of Einstein's theory:
   `relativity of simultaneity' and `length contraction' are incorrect, without having
   to even encounter a `light signal' of the type crucial for Einstein's analysis
  of space and time.  In the present section it is demonstrated that the properties
   of light described in Einstein's second postulate are in fact necessary consequences of the 
   LT and the physical nature of light, which consists of massless particles ---photons.
  \par In the approach of the present paper there is, so far, no connection between the limiting
   relative velocity, $V$, introduced in the derivations of the LT presented
   in the Appendix and the speed of light, $c$. In order to make this connection it is 
   necessary to consider the relativistic kinematics of some arbitary physical object of
    Newtonian mass, $m$.
   This is done by introducing the energy-momentum 4-vector, $P$, in the frame S  of an object
    of Newtonian mass $m$, at rest in the frame S', according to
   the definitions:
    \begin{equation}
    P \equiv (P_t,P_x) \equiv  mU \equiv m(\gamma V,\gamma v)
    \end{equation}
   where
     \begin{equation}
  U \equiv \frac{d X}{d t'} = (V\frac{d \tau}{ d t'},  \frac{d \tau }{d t'}\frac{d x}{d \tau})
         = (\gamma V,\gamma v)
    \end{equation}
    and where the differental form $d \tau = \gamma d t'$ of the time dilation relation (5.23) has been
    used. 
   Here $X \equiv (V\tau,x)$ is the space-time 4-vector specifying the position of the 
     object in S, $t'$ is the apparent time in S of a clock in the rest frame, S', of the object,
    and $U$ is its 4-vector velocity.
   The relativistic energy, $E$,  and momentum,
   $p$, of the object are then defined as:
   \begin{eqnarray}
   E \equiv V P_t & = & m \gamma V^2, \\
   p \equiv P_x & = & m \gamma v.
 \end{eqnarray}
    These last two equations together with the definition of the parameter $\gamma$ after
   Eq.~(5.18) above, lead to the important kinematical relations connecting $m$, $V$, $p$ and $E$:
   \begin{eqnarray}  
   E^2 & = & m^2 V^4 +p^2 V^2, \\
   v & = & \frac{p V^2}{E} = \frac{p V^2}{(m^2 V^4 +p^2 V^2)^\frac{1}{2}}.
 \end{eqnarray}
  It is an immediate consequence of Eq.~(7.6) that the speed of a massless physical 
  object, in any inertial frame, is just $V$. If light consists of such massless particles
  it follows that $c = V =~\rm{constant}$. This is the first part of Einstein's second
  postulate.
  \par To prove the second part of the second postulate: that the speed of light does
  not depend on that of its source, the parallel velocity addition formula, readily derived from
  the LT equations (5.17) and (5.18), may be used: 
   \begin{equation}
 w = \frac{u +v}{1+\frac{u v}{V^2}} =  \frac{u +v}{1+\frac{u v}{c^2}}.
    \end{equation}
   Here, $u = d x'/d \tau'$ is the velocity of an object relative to S' and $w = dx/d \tau$ is the velocity
  of the same object relative to S. Suppose the the light source is a rest in S'. The velocity
  of the source for an observer at rest in S is then $v$, and, since S' is an inertial
   frame, $u = u_{\gamma} = c$. The velocity of the photon as observed in S is then given by
   (7.6) as:
   \begin{equation}
 w_{\gamma} = \frac{c +v}{1+\frac{v}{c}} = c.  
    \end{equation}
   i.e., the speed of the photon in S does not depend on the velocity, $v$, of the source
      in S. This is the second part of the second postulate. It has now been shown
    that the second postulate necessarily follows from the space-time LT and the fact
    that light consists of massless particles.
    \par Einstein had himself discovered, earlier in 1905, that light indeed {\it does
    consist} of massless particles: his `light quanta', work for which he was later awarded
     a Nobel Prize, but did not realise that this 
    together with the space-time LT, derived in his special relativity paper, necessarily
    implied the correctness of
    the second postulate. However, in 1905, the relativistic kinematics embodied in Eq.~s(7.5)
    and (7.6) above, although already implicit in formulae
     given in Ref.~\cite{Ein1}, had not yet been developed.
    This was done by Planck in 1907~\cite{PlanckRK}.
    \par Note that a complete understanding of the second postulate does not
     require the
     introduction of any concept of classical electromagnetism, or any other dynamical theory.
    In spite of the title of 
     Einstein's seminal paper, the fundamental concepts necessary for a complete
    description of the physics of space and time intervals do not depend, in any way, on those
    of classical electrodynamics, or of any other specific domain of physics.
     Since, however, any physical law must, when correctly formulated, respect the space-time
   geometry embodied in the space-time LT, as well as the kinematical relations (7.5) and (7.6),
   the study of such laws can be used to derive the LT, i.e. to reveal the nature of the
    underlying space-time geometry. Thus Einstein's special relativity paper with its strong
   emphasis on classical electrodynamics was actually, and in just the same way as the earlier
   work of Voigt~\cite{Voigt}, Larmor~\cite{Larmor} and Lorentz~\cite{Lor} and the contemporary
  work of Poincar\'{e}~\cite{Poin}, of an essentially
  heuristic nature. Indeed, from the perspective of a later century, much more so, in
   spite of Einstein's assertion, than the light quantum paper.
   \par The incorrect conclusions of Einstein's paper (misinterpretations of the
    physical meaning of the space-time LT) result from an insufficiently rigorous
   discussion of actual measurements of space and time intervals and their relation to 
   clock synchronisation  procedures. More precisely, they were due to a mismatch between
   the interpretation
   given to certain symbols and the actual physical measurements to which they should 
   correspond.

    \SECTION{\bf{Summary and Discussion}}
     The aim of the present paper has been to construct the fundamental physics of 
    space and time (in the absence of gravitational effects as described by the
    general theory of relativity) on a more secure foundation than hitherto. Two aspects
    are discussed: the axiomatic basis of the space-time LT and the experimental science
    of the measurement of space and time intervals.
    \par  After a brief discussion, in the Introduction,
    of some initial postulates, two previously published~\cite{JHF1,JHF2} axiomatic derivations of the
      LT are
     recalled in the Appendix. Two of the three postulates used are weak, and may even seem
     to be `obvious'. None of the postulates refer to electrodynamics or any other
     dynamical physical theory. The essential logic of the first derivation, in which a 
     universal parameter, $V$, with the dimensions of velocity necessarily appears in the course
     of the calculation, 
     was already followed in Ignatowsky's derivation of the LT~\cite{Ignatowsky}
     published in 1910. Einstein's electrodynamics-based 1905 derivation of the LT
     is then seen to be of an essentialy heuristic nature, since the simplest logical
     foundations of space-time geometry, as described by the LT, are quite unrelated 
     to Einstein's postulates.
     \par The remainder of the paper is devoted to an attempt to provide a rigorous
       mathematical description of the measurement of space and time intervals in terms
     of `pointer-mark coincidences' or PMCs. Several authors have suggested that such a concept
   should form the basis for specifying the results of all experiments in physical 
   science~\cite{RBLHM,GBB}, but without developing the idea into a calculus that provides
   a precise definition of raw experimental data and their relation to the mathematical
   symbols expressing any theoretical prediction of this data. A proposition for such 
   a `PMC calculus' is given in Section 2.
   \par Different clock synchronisation procedures are described in Sections 3 and 4. Because
    of the dependence of Einstein's light signal synchronisation procedure on 
    postulated physical properties of light, and its possible subsequent `conventionality'~\cite{Convt}
    new, simple, and `convention free' methods of clock synchronisation are proposed.
    The first, by `pointer transport', is applicable to an arbitary number of clocks in the same inertial
     frame. The second method, by `length transport', is applicable to an arbitary number of clocks 
    in the same, or two different, inertial frames. A practical device, shown in Fig.~6, to synchronise
   pairs of clocks, one in each of two different inertial frames, is proposed. The associated PMC in the
    two frames
    are constructed with the aid of similar light sources, light detectors and screens, at rest in each
   of the frames. 
   \par In Section 5, the relation of measurements of space and time intervals, provided by
     two pairs of, spatially separated, synchronised, clocks in two different inertial frames,
     as predicted by the
     Galilean or Lorentz transformations, is discussed. The length transport method is
     used to mutally synchronise the four clocks. Two spatially separated clocks, C1 and C2, are
     situated in the frame S, while two others,  C1' and C2', with the same spatial
     separation as C1 and C2, are situated in the frame S' moving with a uniform velocity
     relative to S. It is found that the description of the spatial positions of the clocks
    is the same for the GT and the LT. The measured clock separations are the same in S and S' 
    and are therefore Lorentz invariant quantities. Relativistic time dilation, as predicted
    by Einstein, by use of the the LT, is found to exist, but the apparent time is the same for
   {\it all} synchronised clocks at rest in a given inertial frame, when they are viewed by 
    an observer in a different inertial frame ---there is here no `relativity of simultaneity'
    as predicted in Einstein's 1905 paper~\cite{Ein1}. 
    \par In Section 6 it is demonstrated how the spurious `relativity of simultaneity'
     and `length contraction' effects of conventional special relativity theory result from
     an incorrect assignment of space and time offsets in the LT used to describe
     observations of synchronised clocks in different inertial frames.
      As previously pointed out~\cite{JHF3}, the correct description of several
     synchronised clocks in a given reference frame requires the use of a `local'
     LT at each clock, that is, one in which the clock is situated at the origin
     of coordinates in its own proper frame. The experimentally well-verified~\cite{JHF3}
     time dilation effect is predicted by the use of just such a local LT.
   
     \par In Section 7, Einstein's counter-intuitive second postulate stating the constancy
     of the speed of light in different inertial frames, and its source velocity independence,
      is demonstrated to be
      a necessary kinematical consequence of the LT and the identification of light with
      massless particles --photons. This implies that $c = V$ where $c$ is the speed of 
      light and $V$ is the maximum relative velocity of two inertial frames whose
      space and time coordinates are related by the LT.
      \par Further consequences of the restriction, in special relativity, to local LT,
       have been previously discussed in Reference~\cite{JHF3}. The relativistic kinematics
       of point-like physical objects\footnote{This is also true of an extended object whose
       position is specified as that of its centroid.} is 
       the same as in conventional SR, since the position of such objects is typically used as their proper
       frame origin, corresponding to use of a local LT. 
       \par The absence of `relativistic length contraction' indicates that certain calculations
       in classical electrodynamics, where the existence of such an effect is assumed, are in
       need of revision. One example is a derivation of the Heaviside formula for the electric 
      field of a uniformly moving charge~\cite{JackHF,LLHF}, another is the calculation
      of the force between long, parallel, conductors in motion~\cite{FeynCEM,WGVR,EMPEM}.
      See Refs.\cite{JHFRCED,JHFRSKO,JHFIND,JHFSTF,JHFRETF,JHFQEDEXP}. 
   \par Although it has been argued in the present paper and Reference~\cite{JHF3} that the 
      `relativity of simultaneity' effect predicted by Einstein is spurious and unphysical
        its existence (or absence)
        can, unlike the `length contraction' effect, be readily established using modern
      experimental techniques. This is because `relativity of simultaneity'  is an O($\beta$)
      effect whereas `length contraction' is of O($\beta^2$). Satellite experiments to perform
      such a test are proposed in Reference~\cite{JHF5}. In Reference~\cite{JHF3} is 
     proposed a different satellite experiment to test for the existence of an observable  O($\beta^2$)
      effect `time contraction'~\cite{JHF6}, which is absent when special relativistic 
      predictions are based solely on the use of a local space-time LT.

 \newpage

{\bf Appendix}
\renewcommand{\theequation}{A.\arabic{equation}}
\setcounter{equation}{0}
  \par In the following, the main features of two previously published~\cite{JHF1,JHF2}
 derivations of the space-time Lorentz transformation are described. The derivations
 are based on the three postulates stated in the Introduction: (A) Uniqueness, (B) the measurement
  reciprocity postulate and (C) Space-time exchange invariance symmetry. The first derivation~\cite{JHF1}
 is based on postulates (A) and (B), the second~\cite{JHF2} on (A) and (C). As throughout the present paper, for simplicity,
   only space-time events lying along the $x$, $x'$ axes   of the frames S,S', previously introduced, are considered.
  The generalisation of the first derivation to events in three spatial dimensions can be found in~\cite{JHF1}.
 \par Both derivations consider first the transformation of space coordinates between S and S'.
    With a suitable choice of coordinate origins and time offsets, the transformation may be written, in 
    general, as:
    \begin{equation}
     f_x(x',x,t,v) = 0.
    \end{equation}
    The postulate (A) will be satisfied provided that the function $f_x$ in (A.1) has a multi-linear structure:  
   \begin{equation}
       x'+a_1 x + a_2 t +b_1 x x'+b_2 x t +b_3 x' t  +c x x' t = 0
 \end{equation}
    where the coefficients $a_i$, $b_j$ and $c$ may be functions of $v$ but are independent
     of $x'$, $x$ and $t$.
   The velocity of S' relative to S is:
    \begin{equation}
   v \equiv \left. \frac{d x}{d t}\right|_{x' = \chi'}
 \end{equation}
    where $\chi'$ may take any constant value. Differentiating (A.2) with respect to $t$ and using
    (A.3) gives:
    \begin{equation}
     v = - \frac{a_2+ b_2 x + b_3 \chi'+ c \chi' x }{a_1+ b_1 \chi' + b_2'+ c \chi' t }.
 \end{equation}
     Since this equation must hold for all values of $x$ and $t$ it follows that 
      $ b_2 = c = 0$ so that:
    \begin{equation}
     v = -\frac{a_2}{a_1}\frac{(1+\frac{b_3 \chi'}{a_2})}{(1+\frac{b_1 \chi'}{a_1})}.
 \end{equation}
      This equation holds for all values of $\chi'$ provided that:
    \begin{equation}
     \frac{b_3}{a_2} =  \frac{b_1}{a_1}
 \end{equation}
      in which case:
    \begin{equation}
     v = -\frac{a_2}{a_1}.
 \end{equation}
      Substituting the above values of the coefficients into (A.2) gives:
    \begin{equation}
      x' = \frac{a_1(v)(x-vt)}{\frac{b_1(v)(x-vt)}{a_1(v)}-1}.
 \end{equation}
   Making the substitutions $x \rightarrow -x$,  $x' \rightarrow -x'$,  $v \rightarrow -v$,
    i.e. inverting the direction of the $x$,$x'$ axes, the relation 
    \begin{equation}
      x' = \frac{a_1(-v)(x-vt)}{-\frac{b_1(-v)(x-vt)}{a_1(-v)}-1}
 \end{equation}
    is obtained. Consistency with (A.8) requires that:
  \begin{equation}
      a_1(-v) =  a_1(v),~~~~b_1(-v) =  -b_1(v) 
   \end{equation}
  so that $a_1$ is an even, $b_1$ an odd, function of $v$. However  $b_1$ has dimension
   $[{\rm L}^{-1}]$, and since the only quantities in the problem with this dimension are $1/x$ and
   $1/x'$, the definitions of the coefficients in (A.2) and dimensional consistency
   require that $b_1$ vanishes. In this case (A.8) becomes:
    \begin{equation}
     x' = \gamma(x-vt)
 \end{equation}
      where 
    \begin{equation}
     \gamma \equiv -a_1
 \end{equation}
      and $\gamma(v = 0) = 1$.
  \par Considering now the time transformation equation with the same choice of coordinate origins 
       and time offsets as (A.1):    
   \begin{equation}
     f_t(t',x,t,v) = 0.
    \end{equation} 
    The postulate (A) will be satisfied provided that
  \begin{equation}
       t'+A_1 x + A_2 t +B_1 x t'+B_2 x t +B_3 t' t  +C x t' t = 0.
 \end{equation}
  Differentiating with respect to $t$, using the definition (A.3) and rearranging gives:
  \begin{equation}
   v = -\frac{\left[(dt'/dt)[1+B_1x+(B_3+Cx)t]+A_2+B_2x +(B_3+Cx)t'\right]}
             {A_1+B_1t'+(B_2+Ct')t}.
 \end{equation}
   Since $v$ is independent of $x$, $t$ and $t'$,
   \begin{equation}
      B_1 = B_2 = B_3 = C = 0
  \end{equation}
   so that (A.15) may be written as  
\begin{equation}
   v  = -\frac{\left[dt'/dt+A_2\right]}{A_1}.
 \end{equation}
  Using the conditions (A.16) and (A.17) the time transformation equation simplifies to
\begin{equation}
   t' = \gamma^{\dagger}(t-\frac{\delta x}{v})
 \end{equation}
  where
 \begin{eqnarray}
  \gamma^{\dagger} & \equiv & -A_2, \\
   \delta & \equiv & \frac{\tilde{\gamma}\gamma^{\dagger}-1}{\tilde{\gamma}\gamma^{\dagger}}, \\
   \frac{1}{\tilde{\gamma}} & \equiv & \frac{dt'}{dt}.
  \end{eqnarray}
  Solving (A.11) and (A.18), the inverse transformation equations are:
  \begin{eqnarray}
   x & = & \frac{1}{1-\delta}\left[\frac{x'}{\gamma} + \frac{v t'}{\gamma^{\dagger}}\right], \\
   t & = & \frac{1}{1-\delta}\left[\frac{t'}{\gamma^{\dagger}} + \frac{\delta x'}{\gamma v}\right.]
 \end{eqnarray}
 which may be compared with the same equations given by exchanging primed and unprimed quantities and
   setting $v$ equal to $-v$ in (A.11) and (A.18):
   \begin{eqnarray}
   x & = & \gamma(x'+ v t'), \\
   t & = & \gamma^{\dagger}(t'+ \frac{\delta x'}{v}).
 \end{eqnarray} 
 Consistency of (A.22) with (A.24) and of (A.23) with (A.25) requires
 \begin{eqnarray}
   \gamma & = & \frac{1}{\gamma(1-\delta)},~~~~ \gamma v  =  \frac{v}{\gamma^{\dagger}(1-\delta)}, \\
   \gamma^{\dagger} & = & \frac{1}{\gamma^{\dagger}(1-\delta)},~~~~\frac{\gamma^{\dagger}\delta}{v} 
    = \frac{\delta}{\gamma v (1-\delta)}
  \end{eqnarray} 
 so that $\gamma^{\dagger}\gamma = 1/ (1-\delta) =  \gamma^2 = (\gamma^{\dagger})^2$, or
   $\gamma^{\dagger} = \gamma$. Each equation in (A.26) and (A.27) then gives 
 \begin{equation}
  \delta = \frac{\gamma^2 -1}{\gamma^2}
\end{equation}
 and the time transformation equation is written as\footnote{In Ref.~\cite{JHF1}, (A.29) was derived
  directly from the space transformation equation using an argument that did not correctly 
  distinguish a transformation and its inverse from the different transformations 
  describing a primary experiment and its reciprocal, as discussed in Section 5. above. Also the 
  reciprocity relation (A.32) was derived in Ref.~\cite{JHF1} by considering the spurious `length 
   contraction' effect rather than, as here, time dilation.}
 \begin{equation}
   t' = \gamma(t-\frac{\delta x}{v}).
 \end{equation}
 \par Considering now a clock at rest at $x' = 0$ and using (A.11) to eliminate $x$ from (A.29) 
  gives the time dilation relation:
  \begin{equation}
   t' = \frac{t}{\gamma}
 \end{equation}  
   from which it follows that $\tilde{\gamma} = \gamma$. Applying postulate (B) to the reciprocal 
   time dilation experiment gives
  \begin{equation}
    \gamma'(v') =  \gamma(v') =  \gamma(v)
  \end{equation}
    since $\gamma'$ is the same function of $v'$ as  $\gamma$ is of $v$.
   (A.31) requires that
   \begin{equation}
   v' \equiv -\left.\frac{d x'}{d t'}\right|_{x = \chi} =  v \equiv \left. \frac{d x}{d t}\right|_{x' = \chi'}.
 \end{equation} 
     Thus in the reciprocal experiment the velocity of an object at rest in the frame S relative
    to the frame S' is equal and opposite to that of an object at rest in S' relative to 
    the frame S in the primary experiment, described, for example, by (A.11) and (A.29) with
     $x' = 0$\footnote{Notice that (A.32) is essentially a {\it definition} of the reciprocal
     experiment as that in which $v' = v$, rather than a statement about what would be observed in
    the frames S and S' in the primary experiment. The relation (A.32) mistakenly interpreted,
    following Einstein~\cite{Ein1}, in the latter manner, is usually called the `Reciprocity Principle'
    ~\cite{BG} in the literature and text books.}. The reciprocity relation (A.32) is now used to 
    complete the derivation of the LT.
  \par  The parameter $\gamma$ is a even function of $v$, the relative velocity of the frames S and S',
        such that $\gamma(v = 0) =1$. There are two possibilities when $v \ne 0$, $\gamma(v) \ge 1$ or
     $\gamma(v) < 1$. In the former case it follows from (A.28) that:
     the function $\delta(v)$ satisfies the inequality:
    \begin{equation}
     1 \ge\delta(v) \ge 0
 \end{equation}
     and in the latter case the inequality:
    \begin{equation}
     0 > \delta(v) \ge -1.
 \end{equation}
      Consider now the velocity $V$, defined as the solution of the equation $\delta(V) = 1$ 
      when $\gamma(v) \ge 1$ and of the equation $\delta(V) = -1$ when $\gamma(v) < 1$. 
      It is clear from the inequalities (A.33) and (A.34), and that in the former (latter)
      case $\delta(v)$ is an increasing (decreasing) function of $v$, that the meaning of $V$
       is the maximum 
       mathematically allowed relative velocity of any two inertial frames. This is the point
      at which the universal constant, $V$, with dimensions of velocity, makes its entrance into
      a physical world where space and time measurements in different inertial frames are connected
       by the Lorentz transformation.
      \par Consider now a massive physical object moving with velocity $u$ parallel to the positive
      $x'$ axis in S'. Its velocity, $w$, relative to the positive x-axis in S may be derived
      by differentiating (A.11) and (A.29), taking the ratio of the derivatives,
      making use of the definitions of $u$ and $w$:
    \begin{equation}
      u \equiv \frac{d x'}{d t'},
 \end{equation}
    \begin{equation}
     w \equiv \frac{d x}{d t}
 \end{equation}
       and rearranging the equation so obtained to give:
    \begin{equation}
       w = \frac{u +v}{1+\frac{u \delta(v)}{v}}.
 \end{equation}
       Denoting the frame S by A, S' by B and the proper frame of the physical object introduced above
        by C,
       the velocities $v$, $u$ and $w$ may be written, in an obvious notation, as
        relative velocities:
     \[  v = v_{AB},~~~~~u = v_{BC},~~~~~w = v_{AC}. \]
        So that (A.37) may be written:
    \begin{equation} 
  v_{AC}  = \frac{ v_{BC} + v_{AB}}{1+\frac{ v_{BC} \delta(v_{AB})}{v_{AB}}}.
 \end{equation}
      Exchanging the labels A and C in (A.38) gives
    \begin{equation}
   v_{CA}  = \frac{ v_{BA} + v_{CB}}{1+\frac{ v_{BA} \delta(v_{CB})}{v_{CB}}}.
 \end{equation}
    The reciprocity relation (A.32) implies that:
    \[   v_{CA} = -v_{AC},~~~~~v_{BA} = -v_{AB},~~~~~ v_{CB} = -v_{BC}. \]
    Making these substitutions in (A.39) gives:
    \begin{equation}
   v_{AC}  = \frac{v_{BC} + v_{AB}}{1+\frac{ v_{AB} \delta(-v_{BC})}{v_{BC}}}.
 \end{equation}
    Consistency with (A.38) then requires that
    \begin{equation}
    \frac{ v_{BC} \delta(v_{AB})}{v_{AB}} = \frac{ v_{AB} \delta(-v_{BC})}{v_{BC}}.
  \end{equation}
     which in turn requires (restoring the original notation for the three velocities):
    \begin{equation}
   \frac{v^2}{\delta(v)} = \frac{u^2}{\delta(-u)} = \frac{V^2}{\delta(V)}
     = \pm V^2 = {\rm constant}.
 \end{equation}
     The constant in (A.42) is plus or minus the square of the limiting relative velocity
     $V$ discussed after Eq.~s(A.33) and (A.34) above. The plus sign correponds to $\gamma(v) \ge 1$,
     the minus sign to $\gamma(v) < 1$.        
     Since $v$ and $u$ in equation (A.42) are
     independent quantites the ratios in the first two members of (A.42) must both be equal to
     the same constant.\footnote{The mathematical situation here is very similar to that
     of a factorisable solution to a partial differential equation. For example, the Schr\"{o}dinger
     equation for an atom with potential depending only on the radial coordinate. The constants
     relating, in a similar fashion to (A.42), independent terms containing the polar and azimuthal angles
     and the radial coordinate lead to the conserved angular momentum quantum numbers.}
     It is demonstrated in Reference~\cite{JHF1} that the choice of a negative constant results in
     transformation equations
      that do not respect the Uniqueness postulate (A) and must therefore be discarded in the
      present derivation. The existence of the universal constant in (A.42) was pointed 
     out already in 1910 by Ignatowsky~\cite{Ignatowsky}. Note that setting $u = v$ in (A.42)
         implies that $\delta(v) = \delta(-v)$, i.e., $\delta(v)$ is an even function of $v$.
    Writing $\beta \equiv v/V$ enables the equation given by the first and fourth members of (A.42)
      to be written, on taking the plus sign, as $\delta(v) = \beta^2$. Substituting this value of
     $\delta$ in (A.28) and solving
       for $\gamma$ gives:
    \begin{equation}
     \gamma(\beta)_{\pm} = \pm \frac{1}{\sqrt{1-\beta^2}}.
 \end{equation}
        Since (A.11) requires that $x' \rightarrow x$ when $v \rightarrow 0$, $\gamma(0)$ must
       be positive so that the plus sign must be taken in (A.43). The transformation 
     equations relating $x'$ and $t'$ to $x$ and $t$ are then given by (A.11) and (A.29) as
    \begin{equation}
     x' = \gamma(\beta)(x-vt),
 \end{equation}
    \begin{equation}
     t'= \gamma(\beta)(t-\frac{vx}{V^2})
 \end{equation}
     where  
    \begin{equation}
  \gamma(\beta) \equiv  \gamma(\beta)_{+}=  \frac{1}{\sqrt{1-\beta^2}}.
 \end{equation}
     This completes the first derivation of the space-time Lorentz transformation.
     \par The second derivation~\cite{JHF2} follows the same steps as above up to Eq.~(A.11).
        This is the general expression for the space transformation following from the
        postulate (A). As previously, the parameter $\gamma$ in (A.9) is (see the first equation in (A.10))
      some even function of $v$: 
     \begin{equation}
     \gamma(v) = \gamma(-v).
 \end{equation}
      In order to now apply the constraints of Postulate (C), space-time exchange invariance,
       to (A.11) it must first be rendered dimensionally homogeneous. This may be achieved
     by multiplying the time $t$ by a universal constant $V$ with the dimensions of velocity.
     It is clear that this operation may be applied to any quantity with the
     dimensions of time in any equation of physics. Compensating dimensional changes
     can always be made in any physical constants appearing in the equation. In this way all
    space and time coordinates will have the same dimension, [L], and are automatically 
     converted to components of a 4-vector with this dimension.
     In this way the zeroth components of space-time 4-vectors are defined as:
    \begin{equation}
      x^0 \equiv V t,~~~~~ x'^0 \equiv V t'.
 \end{equation}
       This enables (A.11) to be written as:
    \begin{equation}
      x' = \gamma(\beta)(x-\beta x^0).
 \end{equation}
      Applying the space-time exchange operations~\cite{JHF2} $x \leftrightarrow x^0$ and
     $x' \leftrightarrow x'^0$ to (A.49) gives immediately the time transformation equation:
    \begin{equation}
      x'^0 = \gamma(\beta)(x^0-\beta x).
 \end{equation}
     In general the transformation inverse to (A.50) may be written as:
    \begin{equation}
   x^0 = \gamma(\beta')(x'^0+\beta' x').
 \end{equation}   
     The same inverse transformation may be obtained by eliminating $x$ between (A.49) and (A.50):
    \begin{equation}
    x^0 = \frac{1}{\gamma(\beta)(1-\beta^2)}(x'^0+\beta x').
 \end{equation}
     Consistency of (A.51) and (A.52) then requires that:
    \begin{equation}
   \gamma(\beta') = \frac{1}{\gamma(\beta)(1-\beta^2)}
 \end{equation}
  and 
           \begin{equation}
      \beta' = \beta.
 \end{equation}
     (A.53),(A.54) and the condition that $\gamma(\beta) \rightarrow 1$ as $\beta \rightarrow 0$
     then gives (A.46) above. On using (A.48) to  restore $t$ and  $t'$ in (A.49) and (A.50) the 
     Lorentz transformation  (A.44)-(A.46) is obtained. 
     \par It is interesting to note that, in the second derivation, the postulates (A) and (C)
      necessarily imply the validity of the reciprocity relation (A.54). In the first derivation
      it is found to be, instead, a necessary consequence of postulates
      (A) and (B). 
      \par In the above derivations, concerning only events lying along the x, x' axes, no
       assumption concerning the geometrical properties of space-time, or group-theoretical
       properties of the transformation, were needed to derive the
       Lorentz transformation. In the case of events with non-vanishing y and z coordinates
    the additional postulate of spatial isotropy~\cite{JHF1} is required to derive the
    transformation equations of the remaining spatial coordinates:
    \begin{equation}
         y' =y,
 \end{equation}
    \begin{equation}
         z' =z.
 \end{equation}
    The derivation of these equations in Reference~\cite{JHF1} is very similar to that of Einstein
    in Reference~\cite{Ein1}.
\pagebreak


\begin{thebibliography}{99}

\bibitem{Voigt}
 W.Voigt, Goett. Nachr. 1887 p.41. 
 \bibitem{Larmor}
J.Larmor, `Aether and Matter', (C.U.P., Cambridge, 1900) Chapter X1.
\bibitem{Lor}
H.A.Lorentz, Proc. K. Ak. Amsterdam {\bf6} 809 (1904).
 \bibitem{CK}
 C.Kittel, Am. J. Phys. {\bf 42} 726 (1974).
\bibitem{Ein1}
A.Einstein, Annalen der Physik {\bf17}, 891 (1905).
  English translation by W.Perrett and G.B.Jeffery in `The Principle of Relativity'
  (Dover, New York, 1952) p.37, or in `Einstein's Miraculous Year'
  (Princeton University Press, Princeton, New Jersey, 1998) p.123.
\bibitem{JHF1}
J.H.Field, Helv. Phys. Acta.  {\bf 70} 542 (1997);
 arXiv pre-print: http://xxx.lanl.gov/abs/physics/0410262v1; Cited 27 Oct 2004. 
\bibitem{JHF2}
J.H.Field,  Am. J. Phys. {\bf 69}, 569  (2001);
arXiv pre-print: http://arxiv.org/abs/physics/0012011v2; Cited 10 Jan 2001.
\bibitem{JHF3}
J.H.Field, 'The Local Space-Time Lorentz Transformation: a New Formulation
 of Special Relativity Compatible with Translational Invariance',
  arXiv pre-print: http://xxx.lanl.gov/abs/physics/0501043. Cited 30 Nov 2007.
\bibitem{JHFUMC}
J.H.Field, `Spatially-separated synchronised clocks in the same inertial frame:
  Time dilatation, but no relativity of simultaneity or length contraction´, 
 arXiv pre-print: http://xxx.lanl.gov/abs/0802.3298; Cited 4 Mar 2008.
\bibitem{JHFCRCS}
J.H.Field, `Clock rates, clock settings and the physics of the
  space-time Lorentz transformation',
  arXiv pre-print: http://xxx.lanl.gov/abs/physics/0606101, Cited 4 Dec 2007.
\bibitem{JHFACOORD}
J.H.Field, 'Translational invariance and the space-time Lorentz
  transformation with arbitary spatial coordinates',
arXiv pre-print: http://xxx.lanl.gov/abs/physics/0703185v2, Cited 15 Feb 2008. 
\bibitem{Convt}
See. for example, P.Anderson, I.Vetharaniam and G.E.Stedman,
 Physics Reports {\bf 295} 93 (1998) and references therein.
\bibitem{BG}
 V.Berzi and V.Gorini, Journ. Math. Phys. {\bf 10} 1518 (1969).
\bibitem{Gal}
 Galileo Galilei, `Dialogue Concerning the Two Chief World Systems'
 English translation, Ed J.A.Green (Greenwood Research, Wichita, Kansas, 2000) p.186.
\bibitem{Ignatowsky}
 W.Ignatowsky, Phys. Zeitschr. {\bf 11} 972 (1910).
\bibitem{Pauli}
  W.Pauli, `Relativit\"{a}tstheorie' (Springer, Berlin 2000). English
  translation,`Theory of Relativity' (Pergamon Press, Oxford, 1958) Section 4, p.11.
 \bibitem{LK}
 A.R.Lee and T.M.Kalotas, Am. J. Phys. {\bf 43}, 434  (1975).
  \bibitem{LLB}
 J.M.L\'{e}vy-Leblond, Am. J. Phys. {\bf 44}, 271  (1976).
  \bibitem{Mermin}
 N.D.Mermin, Am. J. Phys. {\bf 52}, 119  (1984).
\bibitem{AS}
 A.Sen, Am. J. Phys. {\bf 62}, 157  (1994).
\bibitem{Poin}
 H.Poincar\'{e}, Rend. del Circ. Mat. di Palermo, {\bf 21} 129 (1906).
\bibitem{RBLHM}
 R.B.Lindsay and H.Margenau, `Foundations of Physics',
 (Dover, New York, 1957) Section 1.1, P5.
 \bibitem{GBB}
 G.Burniston Brown, Proc. Phys. Soc. Lond.{\bf 53} 418 (1941).
 \bibitem{FeynDS}
 R.P.Feynman. `The Character of Physical Law', (M.I.T. Press, Cambridge, Massachusetts 1986)
  pps. 130, 145.
 \bibitem{Ein2}
A.Einstein, Annalen der Physik {\bf 17} 132 (1905).
\bibitem{JHF4}
J.H.Field,  `On the Real and Apparent Positions of Moving Objects in
 Special Relativity: The Rockets-and-String and Pole-and-Barn Paradoxes
 Revisited and a New Paradox',
   arXiv pre-print: http://xxx.lanl.gov/abs/physics/0403094v3. Cited Nov 2007.  
 \bibitem{PlanckRK}
M.Planck, Verh. Deutsch. Phys. Ges.  {\bf 4} 136 (1906).
 \bibitem{JackHF}
J.D.Jackson, `Classical Electrodynamics', 3rd Ed (Wiley, New York, 1998) pps. 558,560.
 \bibitem{LLHF}
L.D.Landau and E.M.Lifshitz, `The Classical Theory of Fields',
   2nd Edition (Pergamon Press, Oxford, 1962) Section 38, p.103.
 \bibitem{FeynCEM}
R.P.Feynman, The Feynman Lectures in Physics, `Electromagnetism I'
   (Addison-Wesley, Reading Massachusetts, 1966) Section 13.6. 
 \bibitem{WGVR}
 W.G.V.Rosser, `Introductory Relativity',
 (Butterworth, London, 1967) Section 7.2, p.220.
 \bibitem{EMPEM}
 E.M.Purcell, `Electricity and Magnetism', Berkeley Physics Course Vol II,
   (McGraw Hill, New York, 1965) Chspter 5.
\bibitem{JHFRCED}
J.H.Field, Phys. Scr. {\bf 74} 702 (2006);
  arXiv pre-print: http://xxx.lanl.gov/abs/physics/0501130v5; Cited 4 Dec 2006
\bibitem{JHFRSKO}
 J.H.Field,  Int. J. Mod. Phys. A  Vol 23 No 2 327 (2008); 
  arXiv pre-print: http://xxx.lanl.gov/abs/physics/0507150v3; Cited 5 Jul 2007 
\bibitem{JHFIND}
 J.H.Field, `Inter-charge forces in relativistic classical electrodynamics:
  electromagnetic induction in different reference frames',
  arXiv pre-print: http://xxx.lanl.gov/abs/physics/0511014v4. Cited 7 Apr 2008
\bibitem{JHFSTF}
 J.H.Field, `Space-time transformation properties of inter-charge forces and
  dipole radiation: Breakdown of the classical field concept in relativistic
  electrodynamics',
    arXiv pre-print: http://xxx.lanl.gov/abs/physics/0604089v4. Cited 4 Apr 2008 
\bibitem{JHFRETF}
  J.H.Field, `Retarded electric and magnetic fields of a moving charge:
 Feynman's derivation of Li\'{e}nard-Wiechert potentials revisited',
 arXiv pre-print: http://xxx.lanl.gov/abs/0704.1574v2. Cited 14 Apr 2008. 
\bibitem{JHFQEDEXP} 
  J.H.Field,  Phys. Part. Nucl. Lett. {\bf 6} 320 (2009);
 arXiv pre-print: http://arxiv.org/abs/0706.1661v3. Cited 17 Jul 2007.
\bibitem{JHF5}
J.H.Field, `Proposals for Two Satellite-Borne Experiments to Test
 Relativity of Simultaneity in Special Relativity',
  arXiv pre-print: http://xxx.lanl.gov/abs/physics/0509213v4. Cited 3 Jun 2011.
\bibitem{JHF6}
J.H.Field, Am. J. Phys. {\bf 68} 267 (2000);
 arXiv pre-print:http://arxiv.org/abs/physics/0004012v1. Cited 8 Apr 2000. 
\end{thebibliography}
\end{document}